\documentclass[a4paper,12pt]{article}
\usepackage[utf8x]{inputenc}
\usepackage{graphicx}
\usepackage{overcite}
\usepackage{amsmath}
\usepackage{amssymb}
\usepackage[T1]{fontenc}
\usepackage{url}

\newcommand{\onlinecite}[1]{\hspace{-1 ex} \nocite{#1}\citenum{#1}}
\newcommand*\diff{\mathop{}\!\text{d}}

\title{\huge \textsf{Mechanism Deduction from\\ Noisy Chemical Reaction Networks}}
\author{Jonny Proppe and Markus Reiher\thanks{\textsf{corresponding author: markus.reiher@phys.chem.ethz.ch}}}

\begin{document}

\maketitle

\vspace*{-0.9cm}\begin{center}
Laboratory of Physical Chemistry, ETH Zürich, \\ Vladimir-Prelog-Weg 2, 8093 Zürich, Switzerland 
\end{center}


\begin{abstract}
\small
	We introduce \textsc{KiNetX}, a fully automated meta-algorithm for the kinetic analysis of complex chemical reaction networks derived from semi-accurate but efficient electronic structure calculations.
	It is designed to (i) accelerate the automated exploration of such networks and (ii) cope with model-inherent errors in electronic structure calculations on elementary reaction steps.
	We developed and implemented \textsc{KiNetX} to possess three features.
	First, \textsc{KiNetX} evaluates the kinetic relevance of every species in a (yet incomplete) reaction network to confine the search for new elementary reaction steps only to those species that are considered possibly relevant.
	Second, \textsc{KiNetX} identifies and eliminates all kinetically irrelevant species and elementary reactions to reduce a complex network graph to a comprehensible mechanism.
	Third, \textsc{KiNetX} estimates the sensitivity of species concentrations toward changes in individual rate constants (derived from relative free energies), which allows us to systematically select the most efficient electronic structure model for each elementary reaction given a predefined accuracy.
	The novelty of \textsc{KiNetX} consists in the rigorous propagation of correlated free-energy uncertainty through all steps of our kinetic analyis.
	To examine the performance of \textsc{KiNetX}, we developed \textsf{AutoNetGen}.
	It semirandomly generates chemistry-mimicking reaction networks by encoding chemical logic into their underlying graph structure.
	\textsf{AutoNetGen} allows us to consider a vast number of distinct chemistry-like scenarios and, hence, to discuss the importance of rigorous uncertainty propagation in a statistical context.
	Our results reveal that \textsc{KiNetX} reliably supports the deduction of product ratios, dominant reaction pathways, and possibly other network properties from semi-accurate electronic structure data.
\end{abstract}


\section{Introduction}
A detailed understanding of reactive chemical systems on arbitrary time scales would support the optimization of chemical processes through directed manipulations of promoting and interfering factors during the course of a reaction. 
	For this purpose, we need to uncover the kinetic properties of complex chemical reaction networks from a first-principles perspective as highly accurate free-energy differences are required for modeling the temporal progress of a reaction.
	However, even electronic structure data may lack the necessary degree of accuracy, despite being derived from the first principles of quantum mechanics.
	Therefore, our objective is to model the kinetics of complex chemical processes on a given time scale with rigorous uncertainty quantification. 
	Such a procedure will ultimately allow us to automatically deduce product distributions, reaction mechanisms, and other network properties from semi-accurate electronic structure data, and to assess their reliability.
	
	There are two steps required prior to the kinetic analysis of a chemical reaction network.
	First, the possibly vast chemical reaction space underlying the problem at hand needs to be explored, at least partially, to generate a  reaction network.
	This preparatory task has been considered in detail by us and others\cite{broadbelt1994, kayala2012, maeda2013, zimmerman2013, rappoport2014, wang2014, bergeler2015, dontgen2015, habershon2015, suleimanov2015, zimmerman2015, dewyer2017, gao2016, habershon2016, wang2016, simm2017a, dontgen2018};
	see also Ref.~\onlinecite{dewyer2018} for a recent review of this topic.
	Any exploration attempt targeting mechanistic completeness is faced with the major challenge to handle a possibly factorial growth of reaction paths for an increasing number of species.
	Obviously, this combinatorial problem will rapidly force every exploration algorithm to stop after a few layers, the number of which may be smaller than what would be needed for a truly comprehensive understanding of a complex  reaction mechanism.
	This unsystematic search (in terms of kinetics) will lead to exploring regions of reaction space that are kinetically irrelevant under given external reaction conditions.
	
	Second, the reaction network under consideration needs to be endowed with parameters (relative free energies or rate constants) obtained from electronic structure calculations including estimates of their correlated uncertainty\cite{frederiksen2004, mortensen2005, petzold2012, wellendorff2012, medford2014, pandey2015, simm2016, proppe2016, pernot2017a, simm2018} (originating from, e.g., approximate exchange--correlation functionals or partition functions).
	As an alternative to ensembles of electronic structure models, advanced machine learning methods (e.g., Gaussian process regression or Bayesian neural network regression)\cite{bishop2009} could be employed for the accurate estimation of uncertainty-equipped model parameters.
	The state of the art in reaction rate theory has recently been presented and discussed at the 2016 Faraday Discussion on Reaction Rate Theory\cite{althorpe2016, althorpe2016a, angulo2016, althorpe2016b}, where we have already presented the principle workflow to arrive at first-principles free energies equipped with correlated uncertainties at the example of a small model network of the formose reaction.\cite{proppe2016}
	We showed that the propagation of correlated uncertainty in activation free energies to time-dependent species concentrations can yield striking variances in equilibration times.
	For a mean standard deviation of about 3 kcal mol$^{-1}$ in the activation free energies (maximum standard deviation of 5.5 kcal mol$^{-1}$), we found a maximum deviation in the equilibration time of almost 23 orders of magnitude.
	
	The focus of this paper is on the kinetic analysis of complex chemical reaction networks.
	Hence, our discussion starts from a network for which the initial species concentrations and all rate constants, including estimates for their correlated uncertainty, are supposed to be known. 
	We define a reaction network endowed with rate constants as the graph representation of a \textit{kinetic model} based on mass action, which is expressed as a system of ordinary differential equations (ODEs) consisting of variables (species concentrations) and parameters (rate constants).
	Since numerical integration of ODE systems is not restricted to chemical kinetics but is relevant in basically all areas of science, there exists a plethora of computer programs for this purpose.
	A review of the corresponding algorithms would go beyond the scope of this paper. 
	We refer to Chapter 9.1 in Ref.~\onlinecite{turanyi2014} for a concise overview of ODE solvers employed in the field of chemical kinetics.

	The number of software packages for kinetic modeling is rich and has a long tradition in chemical kinetics.\cite{turanyi2014}
	For many chemical problems, there exist tailored codes, a personal selection of which is briefly introduced hereafter.
	One of the most widely applied software packages for comprehensive kinetic modeling is CHEMKIN.\cite{kee1980, kee1989, kee1996} 
	As the development of CHEMKIN was and is particularly driven by gas phase chemistry, it comprises application-specific features that take care of, e.g., transport processes or changes in pressure and/or temperature during the course of a reaction.
	An open-source alternative in gas phase chemistry to CHEMKIN with comparable feature scope is Cantera.\cite{goodwin2009}
	Another open-source software package developed for comprehensive kinetic modeling is COPASI (COmplex PAthway SImulator).\cite{hoops2006} 
	Due to its focus on biochemical systems, where particle numbers are likely to be very small, COPASI contains an implementation for stochastic kinetic simulations\cite{gillespie2007} in addition to its deterministic counterpart (numerical integration of ODEs).
	The master equation, which is the fundamental equation for stochastic chemical kinetics, is also crucial in cases where the time scales of reactive events and collisional relaxation compete with each other, such that a nonequilibrium description of state transitions becomes necessary. 
	MESMER (Master Equation Solver for Multi-Energy Well Reactions) has been developed for this purpose by Glowacki and colleagues\cite{glowacki2012, shannon2018} based on their results for both gas phase\cite{glowacki2012a} and solution phase\cite{glowacki2010, goldman2011} chemistry.
	
	To match the philosophy of our developments for a new kind of computational quantum chemistry (SCINE\cite{scine}), we require a comprehensive kinetic network analysis to be based on the following steps: 
\begin{enumerate}
\item[(i)] Translation of a reaction network endowed with uncertainty-equipped rate constants to an ensemble of kinetic models, each model representing a unique set of rate constants;
\item[(ii)] numerical integration of the ensemble of (possibly stiff) kinetic models;
\item[(iii)] identification of possibly relevant species based on noisy concentration data to guide the search for new species;
\item[(iv)] elimination of kinetically irrelevant species and elementary reactions based on noisy concentration data (statistical mechanism deduction);
\item[(v)] global sensitivity analysis to rank reactions according to how much concentration noise the uncertainty of the underlying rate constants induce.
\end{enumerate}

	Here, we introduce \textsc{KiNetX}, a meta-algorithm that accomplishes all of these tasks in a fully automated fashion.
	The two key aspects of \textsc{KiNetX} are that it (i) can steer the exploration of chemical reaction space in order to accelerate this exploration and (ii) enables routine statistico-kinetic analyses of reaction networks endowed with rate constants for which correlated uncertainty information is available.
	This way, we appreciate recent developments related to the exploration of chemical reaction networks\cite{broadbelt1994, kayala2012, maeda2013, zimmerman2013, rappoport2014, wang2014, bergeler2015, dontgen2015, habershon2015, suleimanov2015, zimmerman2015, dewyer2017, gao2016, habershon2016, wang2016, simm2017a, dontgen2018, dewyer2018} and the uncertainty quantification of reaction energies,\cite{frederiksen2004, mortensen2005, petzold2012, wellendorff2012, medford2014, pandey2015, simm2016, proppe2016, pernot2017a, simm2018} both derived from the first principles of quantum mechanics.
	The novelty of \textsc{KiNetX} consists in the uncertainty quantification approach it is built on.
	Instead of making educated guesses, consulting rules, or fitting to experimental data,\cite{turanyi2014} it has become possible to estimate the correlated uncertainty of free-energy predictions from ensembles of electronic structure models, which allows for a direct evaluation of the underlying free-energy covariance matrix.
	The explicit propagation of these correlated uncertainties through the entire workflow renders \textsc{KiNetX} a novel toolbox for statistico-kinetic modeling that significantly increases the reliability of mechanistic conclusions drawn from quantum chemical reaction data.
	We aimed at a very generic input format that does not enforce any context-related formalities.
	Even nonchemical problems (if expressed as mass action-type ODEs) can be studied.
	The uncertainty framework of \textsc{KiNetX} can in principle be coupled to any of the kinetic modeling codes mentioned above if a suitable parser is provided.
	To examine the importance of rigorous uncertainty propagation in kinetic modeling, we developed an automated network generator, \textsf{AutoNetGen}.
	\textsf{AutoNetGen} creates chemistry-mimicking reaction networks based on chemical logic with which we can study an arbitrary number of chemistry-like scenarios.	
	
	This paper is organized as follows:
	In Section \ref{sec:methods}, we introduce the basic equations of kinetic modeling, highlight the importance of uncertainty quantification for this field, and discuss the technical details of \textsc{KiNetX} and \textsf{AutoNetGen}.
	In Section \ref{sec:results}, we present the \textsc{KiNetX} workflow at a specific example and provide statistics on the reliability increase by incorporating uncertainty quantification into the kinetic modeling framework.
	
	
\section{Theoretical and Algorithmic Details}
\label{sec:methods}
\subsection{Kinetic Modeling from a Network Perspective}
	We describe the structure of a chemical reaction network by a graph of $N$ vertices and $L$ bidirectional edges.
	The network is strictly bidirectional as we assume every chemical transformation to be reversible.
	Either of both edges corresponding to a \textit{reaction pair} (a reversible elementary reaction) is assigned an arbitrary but unique direction (\textit{forward},~$+$; or \textit{backward},~$-$).

	We define the $N$-dimensional column vector of time-dependent species concentrations,
\begin{equation}
\mathbf{y}(t) = 
\begin{pmatrix}
y_1(t), & \cdots, & y_N(t)
\end{pmatrix}^\top \ ,
\end{equation}
which keeps track of the population density of each vertex at a given time.
Here, $y_n(t)$ refers to the concentration of the $n$-th chemical species at time $t$, which is a strictly nonnegative quantity.
	The $2L$-dimensional column vector of rate constants,
\begin{equation}
\mathbf{k} = 
\begin{pmatrix}
\mathbf{k}^{+} \\
\mathbf{k}^{-}
\end{pmatrix} =
\begin{pmatrix}
k_1^{+}, & \cdots, & k_L^{+}, & k_1^{-}, & \cdots, & k_L^{-}
\end{pmatrix}^\top \ ,
\end{equation}
contains strictly nonnegative scaling factors that determine the transition rate for each direction of an edge.
	We define the $(N \times L)$-dimensional stoichiometry matrix of forward reactions,
\begin{equation}
\mathbf{S}^{+} =
\begin{pmatrix}
S_{11}^{+} & \cdots & S_{1L}^{+} \\
\vdots & \ddots & \vdots \\
S_{N1}^{+} & \cdots & S_{NL}^{+} 
\end{pmatrix} \ ,
\end{equation}
where the element $S_{nl}^{+}$ ($n \in \lbrace 1,\cdots,N \rbrace$ and $l \in \lbrace 1,\cdots,L \rbrace$) describes the number of molecules of the $n$-th species that is consumed in the $l$-th forward reaction.
	Analogously to $\mathbf{S}^{+}$, we define the $(N \times L)$-dimensional stoichiometry matrix of backward reactions,
\begin{equation}
\mathbf{S}^{-} =
\begin{pmatrix}
S_{11}^{-} & \cdots & S_{1L}^{-} \\
\vdots & \ddots & \vdots \\
S_{N1}^{-} & \cdots & S_{NL}^{-} 
\end{pmatrix} \ .
\end{equation}
	All elements in $\mathbf{S}^+$ and $\mathbf{S}^-$ are strictly nonnegative integers. 
	The combined $(N \times L)$-dimensional stoichiometry matrix reads
\begin{equation}
\mathbf{S} = \mathbf{S}^{-} - \mathbf{S}^{+} \ .
\end{equation}

	From the quantities introduced above and assuming mass action kinetics, we can express the $L$-dimensional column vectors of forward ($+$) reaction rates and backward ($-$) reaction rates as
\begin{equation}
\mathbf{f}^{+/-}(t) = 
\begin{pmatrix}
f_1^{+/-}(t), & \cdots, & f_L^{+/-}(t)
\end{pmatrix}^\top \ ,
\end{equation}
with components
\begin{equation}
\label{eq:f+}
f_l^{+/-}(t) = k_l^{+/-} \prod_{n=1}^N \Big(y_n(t)\Big)^{S_{nl}^{+/-}} \ .
\end{equation}
	By definition, the product of concentrations in Eq.~(\ref{eq:f+}) only equals zero if a species involved in the $l$-th forward/backward reaction (indicated by a positive value of $S_{nl}^{+/-}$) is not populated (zero concentration), since for all species not involved we obtain a factor of 1 due to ${S_{nl}^{+/-} = 0}$.
	Based on the combined stoichiometry matrix, $\mathbf{S}$, and the combined $L$-dimensional column vector of reaction rates,
\begin{equation}
\mathbf{f}(t) = \mathbf{f}^{+}(t) - \mathbf{f}^{-}(t) \ ,
\end{equation}
we can express the time-dependent change in the species concentrations as
\begin{equation}
\mathbf{g}(t) =
\begin{pmatrix}
g_1(t), & \cdots, & g_N(t)
\end{pmatrix}^\top =
\frac{\diff}{\diff t} \mathbf{y}(t) = \mathbf{S} \, \mathbf{f}(t) \ .
\end{equation}
The relevant procedure prior to any kinetic analysis of reaction networks is the numerical integration of $\mathbf{g}(t)$, which is therefore the central quantity in kinetic modeling.

\subsection{Uncertainty Quantification in Kinetic Modeling}
\label{sec:uq}
	The objective of uncertainty quantification in kinetic modeling\cite{turanyi2014} is to assess the accuracy (or bias) and precision (or variance) of concentration profiles obtained from numerical integration of ODE systems.
	To obtain reliable results, it may be necessary to invest great effort in estimating the \textit{correlated} uncertainty of model parameters (free-energy differences or rate constants).
	The mathematical object we are searching for is the \textit{joint probability distribution} of model parameters.\cite{grimmett1992}
	One way to estimate parameter correlation is the backward propagation of uncertainty observed in measured concentration profiles,\cite{kirk2009} which requires knowledge of the underlying mechanism containing all kinetically relevant elementary reaction steps.
	This strategy is clearly appealing for verifying mechanism completeness, but it does not serve our purpose of understanding chemical reactivity from a first-principles perspective.
	
	It has recently been shown that the neglect of parameter correlation in kinetic models can easily lead to false mechanistic conclusions despite being derived from electronic structure calculations.\cite{sutton2016, proppe2016}
	As the parameters of an electronic structure model are (unknown) functions of chemical space, free energies of reaction pathways comprising similar species will not change independently of each other when the value of an electronic structure parameter is changed.
	Fortunately, recent statistical developments in quantum chemistry\cite{frederiksen2004, mortensen2005, petzold2012, wellendorff2012, medford2014, pandey2015, simm2016, proppe2016, pernot2017a, simm2018} enable the estimation of correlated uncertainty in free-energy differences.
	Still, reliably estimating the correlation between free energies is computationally hard as it requires sampling from ensembles of electronic structure models\cite{simm2016, proppe2016} and, hence, repeated first-principles calculations for all species of the chemical system studied (including transition states and, possibly, other structures along the reaction pathway).
	Even if machine learning models were employed, a comprehensive training set based on a vast number of electronic structure calculations would be necessary.\cite{ramakrishnan2014}
	We have already demonstrated the steps required for the propagation of correlated uncertainty in activation free energies for a model network of the formose reaction.\cite{proppe2016}
	Here, we will generalize the procedure for the study of arbitrary chemical systems.
	
	The estimation of uncertainty of a target quantity from the joint probability distribution of model parameters is referred to as \textit{forward} uncertainty quantification\,---\,also known as \textit{uncertainty propagation}.
	The opposite procedure is referred to as \textit{inverse} uncertainty quantification and is applied to calibration problems\cite{pernot2017, proppe2017} (the backward propagation of concentration uncertainty mentioned above belongs to the latter approach).
	Every statistical analysis implemented in \textsc{KiNetX} is based on uncertainty propagation.
	We consider an ensemble of $B+1$ vectors of rate constants that is obtained by drawing $B$ samples from the joint probability distribution of activation free energies with mean ${\mathbb{E}[\mathbf{A}] = \mathbf{A}_0}$ and variance ${\mathbb{V}[\mathbf{A}] =\boldsymbol{\Sigma}_\mathbf{A}}$, which are subsequently mapped to rate constants based on, e.g., Eyring's transition state theory.\cite{eyring1935}
	Each sampled vector of rate constants is labeled $\mathbf{k}_b$ with $b \in \lbrace 1,\cdots,B \rbrace$.
	An additional vector $\mathbf{k}_0$ is directly obtained from $\mathbf{A}_0$.	
	The ensemble of rate constant vectors, $\mathcal{K}_B = \{ \mathbf{k}_0, \mathbf{k}_1,\cdots,\mathbf{k}_B \}$, is the basis for any bottom-up uncertainty analysis in kinetic modeling and is taken into account explicitly by every subalgorithm of \textsc{KiNetX}.
	
	Note that the setup introduced here neglects third- and higher-order moments of the joint probability distribution of activation free energies, which may be a weak assumption for actual reaction networks derived from first principles.
	To avoid these limitations, one can always sample directly from the ensemble of electronic structure models,\cite{simm2016, proppe2016} which requires repeated first-principles calculations and is, therefore, rather inefficient compared to sampling from $\boldsymbol{\Sigma}_\mathbf{A}$.
	Another possibility not yet explored by us is the application of matrix algebra to construct special matrices that simplify expressions for higher-order moments of joint probability distributions (in particular skewness and kurtosis).\cite{meijer2005}

\subsection{Overview of the \textsc{KiNetX} Meta-Algorithm}
	All reaction networks analyzed with \textsc{KiNetX} in this work were generated with \textsf{AutoNetGen} (Section \ref{sec:netgen}).
	Both algorithms are written in \texttt{Matlab}.\cite{matlab2018a}
	For the numerical integration of (generally stiff) ODE systems, we interface to the \textsf{ode15s} module\cite{shampine1997} of \texttt{Matlab}.

	The \textsc{KiNetX} workflow consists of three core algorithms (Sections~\ref{sec:up}--\ref{sec:msa}), all of which take the \textit{correlated} uncertainty in the underlying model parameters (rate constants) explicitly into account:

\begin{enumerate}

\item \textit{Uncertainty propagation}.

 Solve an ensemble of kinetic models derived from a unique reaction network and estimate the kinetic relevance of every species based on the maximum rate of formation (Section \ref{sec:up}).

\item \textit{Network reduction}. Identify and eliminate all kinetically irrelevant vertices and edges of the network by applying a hierarchy of flux analyses resulting in a sparse network that is a comprehensible representation of the underlying reaction mechanism (Section \ref{sec:netred}).

\item \textit{Sensitivity analysis}. Determine the effect of rate constant perturbations on time-dependent species concentrations through an extended version of Morris screening\cite{morris1991} (Section \ref{sec:msa}).

\end{enumerate}

	The minimal input requirements for \textsc{KiNetX} are:
\begin{itemize}
\item A chemical reaction network with $N$ vertices and $2L$ unidirectional edges (provided by \textsf{AutoNetGen} in this work).
\item A set of $N$ initial species concentrations, $\mathbf{y}_0$ (provided by the user).
\item An ensemble of $B+1$ sets of $2L$ rate constants each,
\begin{equation}
\mathcal{K}_B = \{\mathbf{k}_b\}
\end{equation}
with $b = \{0,\cdots,B\}$, which may be derived from an ensemble of electronic structure models based on, e.g., density functional theory\cite{wellendorff2012, pandey2015, simm2016} (provided by \textsf{AutoNetGen} in this work).
\item A maximum reaction time, $t_\text{max}$, representing a practical time scale or a time scale of interest (provided by the user).
\end{itemize}
	At present, we require the input to be provided in SI units. Optional input parameters are: 
\begin{itemize}
\item The maximum tolerable concentration error, $\varepsilon_\mathbf{y}$, between the exact and an approximate solution.
\item A flux threshold, $G_\text{min}$, above which a chemical species will be considered kinetically relevant.
\item The number of log-distributed time points, $U + 1$, between $t = 0$ and $t = t_\text{max}$ at which species concentrations will be evaluated based on cubic spline interpolation.
\item A confidence level, $\gamma$, important for assessing network properties derived from an ensemble of kinetic models.
\item The number of Morris samples, $C$, considered in our global sensitivity analysis. 
\item Absolute and relative concentration error tolerances considered during numerical integration of ODE systems.

\end{itemize}

	Every execution of \textsc{KiNetX} is based on \textit{one} network with a fixed number of vertices and edges.
	Hence, for steering the exploration of reaction networks, \textsc{KiNetX} needs to be executed repeatedly.
	We designed \textsc{KiNetx} to suggest the next exploration step by estimating the kinetic relevance of every species based on the maximum rate of formation (Section~\ref{sec:exploration}).
We will make the \textsc{KiNetx} program available through our webpage (scine.ethz.ch) in 2019.

	Currently, \textsc{KiNetX} is limited to the analysis of homogeneous and isothermal reactive chemical systems in dilute solution, which constitute a major category of chemical systems found in nature and examined in chemical research.
	One prominent subcategory thereof that attracts much attention in fundamental and industrial research is \textit{homogeneous catalysis}, a field that is rather underrepresented in the kinetic modeling literature.\cite{besora2018}
	For the study of dilute systems, collisions involving three or more reactive species can be considered negligible from a statistical point of view.
	For this reason, each element of the forward and backward reaction rate vectors, $\mathbf{f}^{+}$ and $\mathbf{f}^{-}$, is of the form 
\begin{equation}
f_l^{+/-} = k_l^{+/-} y_{n_{l,1}^{+/-}} y_{n_{l,2}^{+/-}} \ ,
\end{equation}
where the vertex index $n$ is determined by the edge index ($l$), the direction ($+$ or $-$) and the (arbitrary) position in the rate equation ($i = 1,2$).
	In case of a unimolecular reaction, we simply set one of the two vertex indices to $n_{l,i}^{+/-} = 0$, denoting a hypothetical null-species with a defined constant value of $y_0 = 1$ independent of the unit of measurement.

\subsection{The \textsc{KiNetX} Workflow}
\subsubsection{Uncertainty Propagation}
\label{sec:up}
	The first step of our \textsc{KiNetX} workflow is the numerical integration of an ensemble of $B+1$ kinetic models, each of them representing a unique set of rate constants.
	Clearly, the user also has the choice to choose ${B = 0}$ (leading to the usual kinetic modeling setup comprising a single numerical integration), but \textsc{KiNetX} is actually designed to analyze an ensemble of kinetic models.

	To assess the magnitude of noise in the $B$ solutions (\textbf{y}-uncertainty) resulting from the ensemble of sets of rate constants (\textbf{k}-uncertainty), we introduce two measures,
\begin{equation}
\label{eq:delta_B}
\delta_B(t_u) = \frac{1}{2B} \sum_{b=1}^B \sum_{n=1}^N \big| y_{n,b}(t_u) - \big\langle y_{n}(t_u) \big\rangle \big|  \ ,
\end{equation}
where
\begin{equation}
\big\langle y_{n}(t_u) \big\rangle = \frac{1}{B} \sum_{b=1}^B y_{n,b}(t_u) 
\end{equation}
is the mean value of concentrations at time $t = t_u$, and
\begin{equation}
\label{eq:Delta_B}
\Delta_B = \frac{1}{t_\text{max}} \sum_{u=1}^U \Big( \delta_{B}(t_{u-1/2}) \Big) \cdot \Big(t_u - t_{u-1}\Big)  \ .
\end{equation}
	The factor $t_\text{max}$ in the denominator of Eq.~(\ref{eq:Delta_B}) equals the sum of time differences given that ${t_0 = 0}$,
\begin{equation}
t_\text{max} = t_U = t_U - t_0 = \sum_{u=1}^U (t_u - t_{u-1}) \ .
\end{equation}
	Furthermore, we define
\begin{equation}
t_{u-1/2} = \frac{1}{2}(t_{u-1} + t_u) \ .
\end{equation}

	According to Eq.~(\ref{eq:delta_B}), $\delta_B(t_u)$ represents the ensemble-averaged \textbf{y}-uncertainty summed over all species at time ${t = t_u}$.
	Here, we focus on $\delta_B(t_\text{max})$ as it measures the variability in the product distribution, a network property of particular interest for synthetic chemists.
	On the contrary, $\Delta_B$ represents a time-averaged \textbf{y}-uncertainty, i.e., the overall variability of species concentrations between $t_0 = 0$ and $t_U = t_\text{max}$.
	The combination of both measures may be valuable for determining the underlying reaction mechanism.
	For instance, if $\delta_B(t_\text{max})$ is close to zero but $\Delta_B$ is rather large, it is likely that we are faced with both or either of the following two scenarios: (i) The sequence of elementary steps is identical or very similar for some \textbf{k}-vectors, but the uncertainty of the activation barrier associated with the rate-determining step is significant; (ii) different \textbf{k}-vectors suggest different routes to the same metastable sink of the reaction network.
	Furthermore, if both $\delta_B(t_\text{max})$ and $\Delta_B$ are below a user-defined concentration error $\varepsilon_\mathbf{y}$, we can safely neglect the ensemble of kinetic models and base all further kinetic analysis on the \textit{nominal} sample represented by $\mathbf{k}_0$ only.

	When applying Eqs.~(\ref{eq:delta_B})+(\ref{eq:Delta_B}), it is important that the number and identity of the time points $t_u$ are the same for all solutions.
	However, it is very unlikely that the time points obtained from an ODE solver are identical for two different $\mathbf{k}$-vectors.
	For this purpose, \textsc{KiNetX} calculates a log-distributed vector of reference time points, ${(t_0, t_1,\cdots,t_U)^\top}$ with ${t_0 = 0}$ and ${t_U = t_\text{max}}$, which is a function of the user-defined parameters $t_\text{max}$ and $U$.
	To obtain species concentrations at the same time points for other $\mathbf{k}$-vectors, \textsc{KiNetX} interpolates the corresponding solutions through cubic spline smoothing.
	Note that spline smoothing is only reasonable if data noise is negligible, a property which one can assess easily in the case of one control variable (here, time).
	To evaluate the reliability of the cubic spline interpolations, we compared them against results from Gaussian process regression.\cite{rasmussen2006}
	Under suitable assumptions, Gaussian process regression yields an optimal trade-off between fitting and smoothness (cubic splines only ensure the latter property).
	Hence, it can be employed to predict concentrations from previously unconsidered values in the time domain (here, the domain between $t_0 = 0$ and $t_U = t_\text{max}$).
	Furthermore, as Gaussian process regression is a strictly Bayesian method, one may obtain reliable uncertainty estimates for each prediction.
	However, as this regression method scales cubically with the number of data points, it is not suitable for repetitive applications (usually, hundreds of regressions would be necessary).
	In all cases studied, we found that both the mean deviation between the two interpolation methods (cubic spline smoothing versus Gaussian process regression) and the mean predictive variance obtained from Gaussian process regression are negligibly small compared to the mean variance of the concentrations themselves (by a factor of $<$10$^{-6}$).
	Hence, data noise is indeed negligible and cubic splines work well for interpolating concentration profiles.

\subsubsection{Network Reduction}
\label{sec:netred}

	We introduce a hierarchy of two reduction algorithms with increasing sophistication (in terms of both rigor and required computing resources), which identify and eliminate kinetically irrelevant species and elementary reactions.
	Initially, we perform a \textit{detailed flux analysis}, followed by a \textit{local barrier analysis} if the former analysis suggests a reduced model resulting in a concentration error that exceeds the user-defined threshold~$\varepsilon_\mathbf{y}$.
	Both algorithms reflect our interpretation of established kinetic modeling concepts.\cite{turanyi2014}

\textbf{Detailed Flux Analysis (DFA).}
	To keep track of the \textit{net} concentration that has passed an edge between $t = 0$ and $t = t_\text{max}$, we integrate the absolute values of $\{f_l(t)\}$ over that time interval,
\begin{equation}
F_l = \int_{t=0}^{t=t_\text{max}} | f_l(t) |  \ \text{d}t \ ,
\end{equation}
	which we define as the \textit{edge flux} corresponding to the $l$-th reaction pair.
	The vector of edge fluxes, $\mathbf{F}$, is the basis for determining the \textit{vertex flux} corresponding to the $n$-th species,
\begin{equation}
G_n = \big(\mathbf{s}_n^+ + \mathbf{s}_n^-\big) \mathbf{F} \ ,
\end{equation}
	where $\mathbf{s}_n^+$ and $\mathbf{s}_n^-$ represent the $n$-th row of $\mathbf{S}^+$ and $\mathbf{S}^-$, respectively.
	Our DFA implementation approximates the edge flux according to
\begin{equation}
F_l \approx F_l^\text{DFA} = \sum_{u=1}^U |f_l(t_{u-1/2})| \cdot (t_u - t_{u-1}) \ .
\end{equation}
	Analogous to the exact expression, the approximate vertex flux reads
\begin{equation}
G_n^\text{DFA} = \big(\mathbf{s}_n^+ + \mathbf{s}_n^-\big) \mathbf{F}^\text{DFA} \ .
\end{equation}
	If $G_n^\text{DFA} < G_\text{min}$, where $G_\text{min}$ is a user-defined threshold, the $n$-th vertex will be removed from the kinetic model, except for the case where the $n$-th vertex represents a reactant.
	All edges that were connected to at least one of the removed vertices will be removed, too.
	After this procedure, the number of vertices and edges should have decreased significantly.
	
	Note that in the case of a closed system, $G_n^\text{DFA}$ is approaching a finite value with increasing time, which renders $G_\text{min}$ an intuitive threshold that resembles a concentration error.
	The assumption behind our DFA is based on this intuitive interpretation: 
	If a reaction channel with a flux smaller than $G_\text{min}$ is removed, the redistribution of this minute amount of flux (i.e., concentration) should yield a concentration error that is comparable to $G_\text{min}$.
	Therefore, we have a means at hand that potentially allows us to control the concentration error introduced by a specific value of $G_\text{min}$.
	Clearly, choosing a smaller value of $G_\text{min}$ will increase the reliability of the DFA-based solution, but also decreases the possible degree of network reduction.
	
	To determine the accuracy loss caused by a DFA, we introduce the measures
\begin{equation}
\label{eq:delta_pq}
\delta_{pq}(t_u) = \frac{1}{2} \sum_{n=1}^N \big| y_{n,p}(t_u) - y_{n,q}(t_u) \big| 
\end{equation}
and
\begin{equation}
\Delta_{pq} = \frac{1}{t_\text{max}} \sum_{u=1}^U \Big( \delta_{pq}(t_{u-1/2}) \Big) \cdot \Big(t_u - t_{u-1}\Big)  \ ,
\end{equation} 
which resemble the control quantities $\delta_B(t_u)$ and $\Delta_B$.
	Here, however, we compare two specific solutions, $\mathbf{y}_p(t)$ and $\mathbf{y}_q(t)$, with each other, one resulting from the \textit{original} network, and the other resulting from the corresponding \textit{sparse} network obtained through our DFA.
	Note that the number of vertices differs for the original and sparse networks.
	Hence, to render Eq.~(\ref{eq:delta_pq}) practical, we define $N$ to be the number of vertices contained in the original network and set all elements in $\mathbf{y}_\text{sparse}(t)$ to zero that refer to vertices not contained in the sparse network.
	In case that all of the ${B + 1}$ kinetic models yield both $\delta_{pq}(t_u)$-values (again, we focus on ${t_u = t_U = t_\text{max}}$) and $\Delta_{pq}$-values that are below a user-defined concentration error $\varepsilon_\mathbf{y}$, \textsc{KiNetX} considers the DFA-based network reduction reliable.
	However, if there is at least one kinetic model that yields $\delta_{pq}(t_\text{max}) > \varepsilon_\mathbf{y}$ or $\Delta_{pq} > \varepsilon_\mathbf{y}$, it depends on the user-defined confidence level $\gamma$ whether the DFA-based network reduction is considered reliable or not.
	We require the confidence level to fulfill the condition $0 \leq \gamma \leq 1$.
	If the $(1 + \gamma)/2$ quantile of $\delta_{pq}(t_\text{max})$ and/or $\Delta_{pq}$ exceeds $\varepsilon_\mathbf{y}$, the DFA-based network reduction is considered unreliable.
	Here, the lowest and highest possible quantiles represent the median and the maximum of both measures over all $B+1$ samples, respectively.

\textbf{Local Barrier Analysis (LBA).}
	There are certain cases in which a DFA-based network reduction is prone to yield unreliable results.
	One example is autocatalysis.
	Imagine the following model reaction network:
	
	2A $\rightleftharpoons$ B (very slow)
	
	B $\rightleftharpoons$ C (fast)
	
	C + A $\rightleftharpoons$ D (fast)
	
	D + A $\rightleftharpoons$ E (fast)
	
	E $\rightleftharpoons$ 2C (fast)
	
	Even though the first reaction step is very slow, it is required to initiate all other reaction steps.
	Without B, neither C nor D nor E can be formed.
	Even though B is obviously a very important species for the reaction mechanism, it will only be relevant at the very beginning of the reation (i.e., until a minute amount of it has been formed).
	Hence, the vertex flux for B, $G_\text{B}$, will be very small, possibly smaller than the user-defined threshold $G_\text{min}$, leading to a false elimination of this species and the second of the five edges.
	
	In this case, $\delta_{pq}(t_\text{max})$ and $\Delta_{pq}$ will clearly exceed $\varepsilon_\mathbf{y}$, and the LBA will start automatically.
	The idea behind our LBA algorithm is fairly simple:
	Set the rate constants of the first reaction pair to zero, which is equivalent to an infinitely high activation barrier or the removal of the corresponding edge.
	Repeat numerical integration and compare the solution to the original one based on $\delta_{pq}(t_\text{max})$ and $\Delta_{pq}$.
	Set the rate constants of the first reaction pair to their original values.
	Repeat the entire procedure for the second, ..., $L$-th reaction pair and for each of the $B+1$ kinetic models.
	
	After this procedure, every reaction pair is associated with $B+1$ values for $\delta_{pq}(t_\text{max})$ and $\Delta_{pq}$.
	If the $(1+c)/2$ quantile of $\delta_{pq}(t_\text{max})$ and/or $\Delta_{pq}$ is smaller than $\varepsilon_\mathbf{y}$, the corresponding reaction pair will be considered kinetically irrelevant and removed from the network.
	All unconnected vertices will be removed, too.
	Subsequently, the resulting $B+1$ reduced kinetic models are integrated and their solutions are compared to the original ones, again based on $\delta_{pq}(t_\text{max})$ and $\Delta_{pq}$.
	If the $(1+c)/2$ quantile of $\delta_{pq}(t_\text{max})$ and/or $\Delta_{pq}$ is smaller than $\varepsilon_\mathbf{y}$, the LBA-based network reduction will be considered reliable.
	It is still possible that $\varepsilon_\mathbf{y}$ is exceeded as the kinetic models we consider are generally nonlinear:
	If the lack of one edge does not alter the solution and the same result is obtained for another edge, it does not imply that the simultaneous lack of both edges leads to the same conclusion.
	If $\varepsilon_\mathbf{y}$ is still exceeded after the LBA-based network reduction, \textsc{KiNetX} will recover the original network and continue its analysis without any reduction.

\subsubsection{Sensitivity Analysis}
\label{sec:msa}
	In the following, we assume that the concentration profiles of the sparse reaction network reveal pronounced uncertainties such that it is difficult to correctly assign product ratios or to suggest a specific reaction mechanism.
	To estimate to which rate constants the concentration profiles are most sensitive, a sensitivity analysis is required.\cite{turanyi2014}
	For this purpose, the rate constants are perturbed to study how such perturbations affect the time-dependent species concentrations.
	
	In a \textit{local} sensitivity analysis, the model parameters are perturbed one by one from their nominal values (here, the elements of $\mathbf{k}_0$).
	While a local sensitivity analysis is straightforward to conduct and computationally feasible (usually, $L$ kinetic simulations are required), it has a significant disadvantage in that it does take into account the correlation between the model parameters (cf.~Section~\ref{sec:uq}).
	Consequently, one may overlook important correlation effects on the uncertainty of concentration profiles.
	Note that our LBA algorithm resembles an extreme variant of a local sensitivity analysis (cf.~Section~\ref{sec:msa}) where each rate constant is perturbed to its minimal value.
	
	In a \textit{global} sensitivity analysis, the correlation between the model parameters is taken into account, but the process requires significantly more computational resources than a local sensitivity analysis.
	Furthermore, the design of a global sensitivity analysis is not as unambiguous as for a local sensitivity analysis, which explains the existence of several approaches that may require $CL$ (Morris screening, where $C$ is an integer usually much smaller than $B$), $B$ (polynomial chaos), or $2BL$ (Sobol's method) numerical integrations.\cite{turanyi2014}

	Morris screening\cite{morris1991} is among the simplest of global sensitivity analyses as it is not designed to induce a mapping between the model parameters and the model solution.
	Instead, its purpose is to categorize the model parameters as either important or unimportant depending on how strongly changes in them affect the model solutions.
	We are particularly interested in this categorization since we aim for a descriptor that informs us about the quality of rate constants obtained from an efficient (\textit{basic}) quantum chemical model.
	If we find that the uncertainty of some species concentrations is too large to derive sensible conclusions about specific network properties, the results of a global sensitivity analysis will support us in identifying the most critical rate constants that require a reevaluation based on a more sophisticated (\textit{benchmark}) quantum chemical model.
	
	The original version of Morris screening does not take into account the joint probability distribution of the model parameters.
	Here, we introduce an extended variant of Morris screening that explicitly considers this information as it is the key element of \textsc{KiNetX}.
	In our implementation of Morris screening, one starts from the nominal sample $\mathbf{k}_0$ and replaces the elements $k_{0,1}^+$ and $k_{0,1}^-$ with the corresponding elements of $\mathbf{k}_1$, $k_{1,1}^+$ and $k_{1,1}^-$.
	For this new vector of rate constants, $\mathbf{k}_{0_1}$, a numerical integration is carried out.
	Subsequently, the elements $k_{0,2}^+$ and $k_{0,2}^-$ of $\mathbf{k}_{0_1}$ are replaced with the elements $k_{1,2}^+$ and $k_{1,2}^-$ of $\mathbf{k}_1$, and a numerical integration based on the new vector $\mathbf{k}_{0_2}$ is carried out.
	This procedure is repeated $L$ times until we arrive at $\mathbf{k}_{0_{L}} = \mathbf{k}_1$, for which numerical integration was already carried out in the first step of the \textsc{KiNetX} workflow (Section \ref{sec:up}).
	The entire procedure is repeated, now by an element-wise change from $\mathbf{k}_1$ to $\mathbf{k}_2$, and generally by an element-wise change from $\mathbf{k}_{b}$ to $\mathbf{k}_{b+1}$ until $b = C - 1$ is reached.
	In the end, we will have generated solutions for another $C(L-1)$ kinetic models in addition to the $B + 1$ solutions obtained for $\mathcal{K}_B$.
	We are interested in the $C(L-1)$ new solutions and the first $C + 1$ of the $B + 1$ solutions obtained previously, amounting to $CL+1$ solutions relevant for our sensitivity analysis.
	
	For each of these solutions, \textsc{KiNetX} determines the $\delta_{pq}(t_\text{max})$ and $\Delta_{pq}$ metrics, where $p$ and $q$ represent two adjacent $\mathbf{k}$-vectors as constructed by our extended version of Morris screening.
	$CL$ values are obtained for each metric, $C$ thereof for every reaction pair.
	We define the sensitivity coefficients $\lbrace z_l^\delta \rbrace$ and $\lbrace z_l^\Delta \rbrace$ as
\begin{equation}
\label{eq:sens_coeff_1}
z_l^\delta = \frac{1}{C}\sum_{b=0}^{C-1} \delta_{b_l,b_{l-1}}(t_\text{max})
\end{equation}
 and
 \begin{equation}
 \label{eq:sens_coeff_2}
z_l^\Delta = \frac{1}{C}\sum_{b=0}^{C-1} \Delta_{b_l,b_{l-1}} \ ,
\end{equation}
respectively.
	The subscript $b_l$ refers to sample $\mathbf{k}_{b_l}$ and we define $\mathbf{k}_{b_0} = \mathbf{k}_b$.
	Note that in Eqs.\ (\ref{eq:sens_coeff_1}) and (\ref{eq:sens_coeff_2}), we do not divide the $\delta_{pq}(t_\text{max})$ and $\Delta_{pq}$ metrics by the corresponding changes in the rate constants, which would be consistent with the usual definition of sensitivity coefficients.
	Instead, we implicitly define the dimension of each sensivity coefficient to be a concentration divided by the conditional standard deviation of the associated rate constant.
	Here, the term \textit{conditional} relates to the consideration of the current values of all other rate constants.
	This way, we obtain a direct measure of the average effect a rate-constant perturbation has on the solution of the corresponding kinetic model.
	We justify this unusual approach by the fact that all perturbations applied originate from actual samples of the underlying joint probability distribution of rate constants.

	Finally, we can set up a ranking of sensitivity coefficients. 
	The larger $z_l^\delta$ and $z_l^\Delta$, the larger the effect of changes in $k_l^{+/-}$ on the concentration profiles.
	With this ranking at hand, one can systematically improve on both the accuracy and precision of the concentration profiles to reliably suggest product distributions (based on $z_l^\delta$) or specific reaction mechanisms (based on $z_l^\Delta$).
	For an actual chemical reaction network derived from first-principles reaction data, one would reevaluate the critical rate constants with a benchmark quantum chemical model.

\subsection{\textsc{KiNetX}-Guided Network Exploration}
\label{sec:exploration}
	In Sections~\ref{sec:up}~to~\ref{sec:msa}, we have introduced the entire \textsc{KiNetX} workflow, which analyzes one network at a time.
	However, given that electronic structure calculations usually require much more computational resources than a kinetic analysis, it is rather impractical to explore all relevant elementary steps prior to a kinetic analysis (not to be mentioned that this strategy may be even unfeasible).
	For this reason, we designed \textsc{KiNetX} to suggest the next exploration step, which may significantly increase the possible exploration depth.
	
	The coupling of kinetic modeling with mechanism generation is well-known in the reaction engineering community.
	It has been introduced by Broadbelt, Green, and co-workers,\cite{susnow1997} and recently revisited by Green and co-workers\cite{han2017} for their \textsc{Reaction Mechanism Generator},\cite{gao2016}
an exploration software originally designed for the study of gas phase reactions (in particular combustion),\cite{song2004} which has recently been extended to understand the kinetics of solution phase chemistry.\cite{jalan2013}
	Here, we follow a similar strategy but additionally take into account the correlated uncertainty in the rate constants.
	Note that the temperatures relevant in the field of combustion are  much higher than what is usual in solution phase chemistry.
	As the thermal rate constant is a decreasing function of temperature in classical rate theories, \textbf{k}-uncertainty is usually neglected in combustion studies.
	This relationship explains why the \textsc{Reaction Mechanism Generator} does not take \textbf{k}-uncertainty into account.
	
	We start from the reactants (species with nonzero initial concentrations) and attempt to find all direct products, i.e., intermediates that are formed by a \textit{single} elementary reaction of the reactants.
	The reactants are considered \textit{active} whereas the direct products are considered \textit{inactive}.
	Only active species are considered in the exploration procedure, i.e., at this point all possible reaction steps have been discovered (according to the exploration algorithm employed).
	To estimate which of the inactive species is potentially the most relevant for the mechanism, we focus on the \textit{formation} rate of each inactive species (indicated by an asterisk)
	
\begin{equation}
^\rightarrow g_{n^\ast}(t_u) = \mathbf{s}_{n^\ast}^+ \mathbf{f}^-(t_u) + \mathbf{s}_{n^\ast}^- \mathbf{f}^+(t_u) \ .
\end{equation}

	Note that $\mathbf{s}_{n^\ast}^{+}$ is multiplied with $\mathbf{f}^{-}(t)$ since in a backward ($-$) reaction, the left-hand-side species ($+$) are formed.
	An equivalent argument holds for the multiplication of $\mathbf{s}_{n^\ast}^{-}$ with $\mathbf{f}^{+}(t)$.
	If a species reveals a very high formation rate at a specific time during the course of the reaction, we assume that this species may be part of relevant reaction channels.\cite{susnow1997}
	Hence, we are interested in the maximum formation rate of each inactive species with respect to all time points $\{ t_u \}$,
 	
\begin{equation}
^\rightarrow g_{n^\ast}^\text{max} = \max \Big\{ {^\rightarrow g_{n^\ast}(t_0)}, ^\rightarrow g_{n^\ast}(t_1), \cdots, ^\rightarrow g_{n^\ast}(t_U) \Big\}
\end{equation}

	The species with the highest value of $^\rightarrow g_{n^\ast}^\text{max}$ will be promoted active.
	Note that if an ensemble of \textbf{k}-vectors is provided, there is more than one maximum formation rate for each inactive species.
	In this case, we rank the inactive species based on a simple statistical model,
\begin{equation}
\eta\big( {^\rightarrow g_{n^\ast}^\text{max}} \big) = \sqrt{ \big\langle {^\rightarrow g_{n^\ast}^\text{max}} \big\rangle^2 + \frac{1}{B-1} \sum_{b=1}^B \Big( {^\rightarrow g_{n^\ast,b}^\text{max}} - \big\langle {^\rightarrow g_{n^\ast}^\text{max}} \big\rangle \Big)^2 } \ ,
\end{equation}
that takes into account both the ensemble average of maximum formation rates,
\begin{equation}
\big\langle {^\rightarrow g_{n^\ast}^\text{max}} \big\rangle = \frac{1}{B} \sum_{b=1}^B  {^\rightarrow g_{n^\ast,b}^\text{max}} \ .
\end{equation}
and the corresponding variance.
	If two species exhibit the same average maximum formation rate but significantly different variances, the high-variance case will be favored as the corresponding species may lead us to potentially more important regions of the reaction space to be explored.
	
	In the original algorithm by Susnow~et~al.\cite{susnow1997}, which is very similar to the one presented here but does not take into account ensembles of kinetic models, the exploration stops if all inactive species reveal values of $^\rightarrow g_{n^\ast}^\text{max}$ that are below a user-defined rate threshold.
	Grenda~et~al.\cite{grenda2003} discussed an important limitation of the rate threshold:
	In certain cases, e.g., where an autocatalytic cycle is an integral part of the reaction mechanism, some species that are of central mechanistic importance may reveal very small maximum formation rates during the exploration procedure.
	In such cases, the rate threshold may need to be chosen very small to achieve mechanistic completeness, which renders it basically impossible to choose a reasonable threshold for a yet unknown reaction.
	 In the most inefficient case, a kinetics-guided exploration would lead to the same reaction network as a nonguided exploration.

\subsection{Chemistry-Mimicking Networks from \textsf{AutoNetGen}}
\label{sec:netgen}
	\textsf{AutoNetGen} generates chemistry-mimicking networks endowed with parameters (both activation free energies and rate constants) in a layer-by-layer fashion.
	The first layer represents the reactants, which need to be specified on input.
	A new layer is formed combinatorially by exploring all possible reactions between all species of the current and previous layers.
	As \textsf{AutoNetGen} generates reaction networks based on abstract rules instead of deriving them from actual chemical representations (e.g., nuclear coordinates for intermediates and transition states), we cannot resort to descriptors identifying reactive sites\cite{bergeler2015} of molecules.
	Instead, we employ random number generators that either enable or disable the formation of an edge between a set of vertices (see the Appendix for more details on this topic).
	In the current version of \textsf{AutoNetGen}, only closed systems (no particle flux into our out of the system between $t = 0$ and $t = t_\text{max}$) are being generated.
	This limitation is introduced here for the sake of simplicity and not for technical reasons; \textsc{KiNetX} is not restricted to these kinds of systems and can also be applied to study open systems. 
	The minimal input requirements for \textsf{AutoNetGen} are (for optional \textsf{AutoNetGen} input, see the Appendix):
	
	\begin{itemize}
	\item The number of samples, $B$, to be drawn from $\boldsymbol{\Sigma}_\mathbf{A}$.
	\item A thermostat temperature, $T$, for the calculation of rate constants.
	\item The average free-energy increase/decrease, $\mu( A^- - A^+)$, for a new intermediate state ($-$) formed by reaction of an intermediate state already present in the network ($+$).
	\item The average difference between the free energies of a left-hand-side intermediate state ($+$) and its corresponding right-hand-side intermediate state ($-$), $\sigma( A^- - A^+)$.
	\item A minimum activation free energy, $\min (A^\ddagger - A^{+/-})$, with respect to the higher-energy intermediate state. 
	\item A maximum activation free energy, $\max (A^\ddagger - A^{+/-})$, with respect to the higher-energy intermediate state. 
	\item The average free-energy uncertainty, $\langle \sigma_A \rangle$ (required for generating an ensemble of kinetic models).
	\item The maximum number of edges, $L_\text{max}$, to be generated.
	\end{itemize}


\section{Results and Discussion}
\label{sec:results}
\subsection{Exemplary \textsc{KiNetX} Workflow}
\label{sec:xmpl}

	In the following, we present one full run of \textsc{KiNetX} applied to a reaction network randomly generated by \textsf{AutoNetGen} consisting of $N = 103$ vertices and $L = 118$ edges, which we term CRN-X.
	The exploration started from two reactants, \textbf{1} and \textbf{2}, with initial concentrations of 1.00~mol~L$^{-1}$ and 0.50~mol~L$^{-1}$, respectively.
	The molecular mass of \textbf{2} is twice as large as the molecular mass of \textbf{1}.
	\textsf{AutoNetGen} generated an ensemble of $B+1=25$ \textbf{k}-vectors sampled from a covariance matrix representing free-energy uncertainties of ($0.5\pm0.1$)~kcal~mol$^{-1}$.
	The resulting uncertainties of the free energies of activation amount to ($1.1\pm0.6$)~kcal~mol$^{-1}$.
	Rate constants were obtained on the basis of Eyring's transition state theory\cite{eyring1935} for a temperature $T = 298.15$~K.
	A detailed list containing all input values submitted to both \textsf{AutoNetGen} and \textsc{KiNetX} can be found in the Appendix.
	
	Fig.~\ref{fig:pre_msa} shows concentration--time plots for all species that exceeded a concentration threshold of 0.01~mol~L$^{-1}$ during the course of reaction.
	We refer to these species as \textit{dominant} species.
	The 25 different solutions draw a diverse picture:
	In some cases, reactant \textbf{1} is fully consumed at the end of the reaction course, and in other cases, more than 50\,\% of the initial concentration remains at $t = t_\text{max}$.
	Also, the potential main product \textbf{33} reveals final concentrations ranging between about 0.3~mol~L$^{-1}$ and 0.9~mol~L$^{-1}$, the observable value of which will, in turn, affect the concentrations of the potential side products \textbf{29}, \textbf{32}, \textbf{34}, \textbf{42}, and \textbf{68}.
	
\begin{figure} [!ht]
\centering
\includegraphics[width=\textwidth]{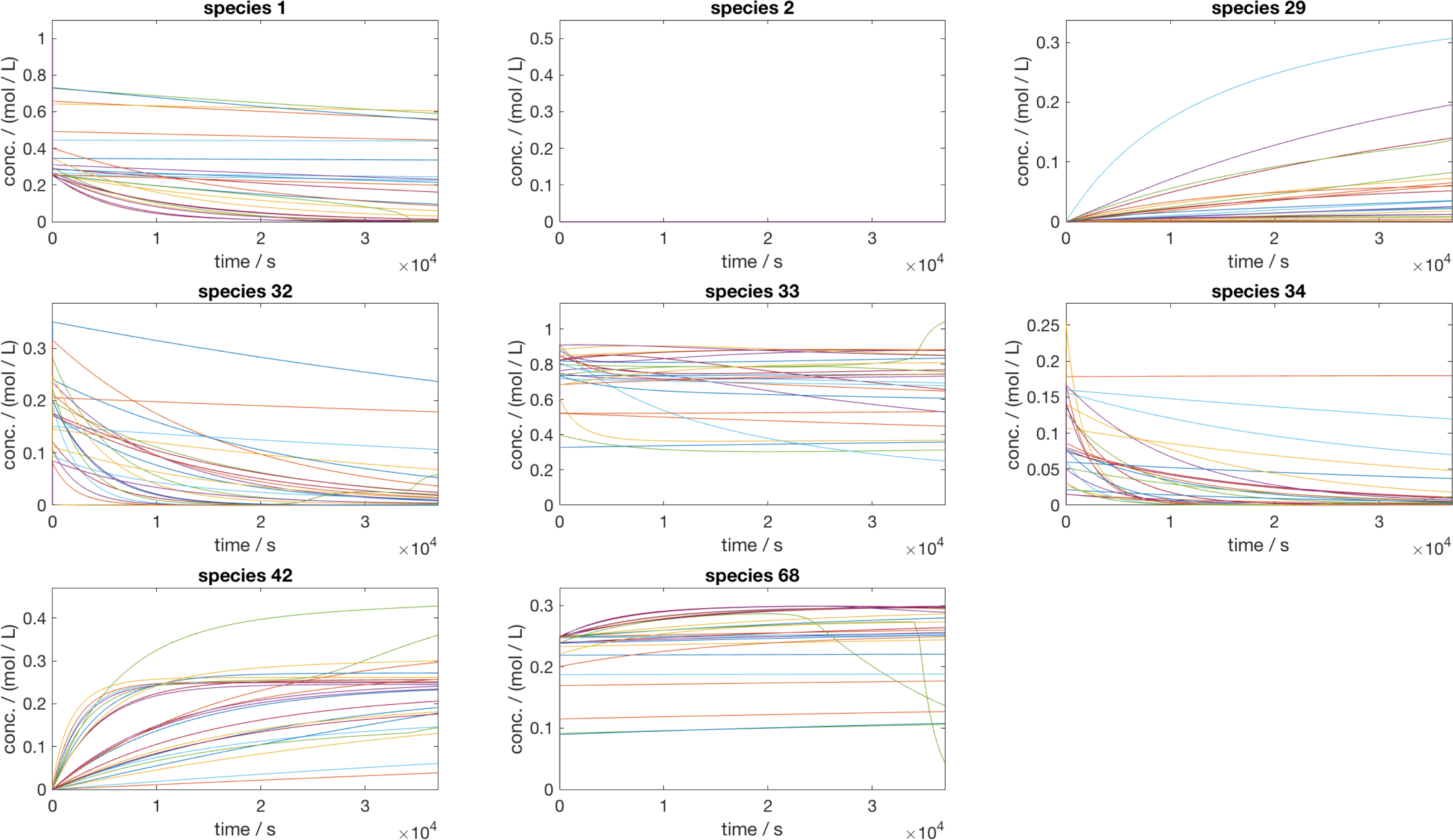}
\caption{
Concentration--time plots for the dominant species of the exemplary reaction network CRN-X \textit{before} any parameter refinement.
An ensemble of 25 distinct kinetic models derived from CRN-X has been considered.
}
\label{fig:pre_msa}
\end{figure}
	
\begin{figure} [!ht]
\centering
\includegraphics[width=\textwidth]{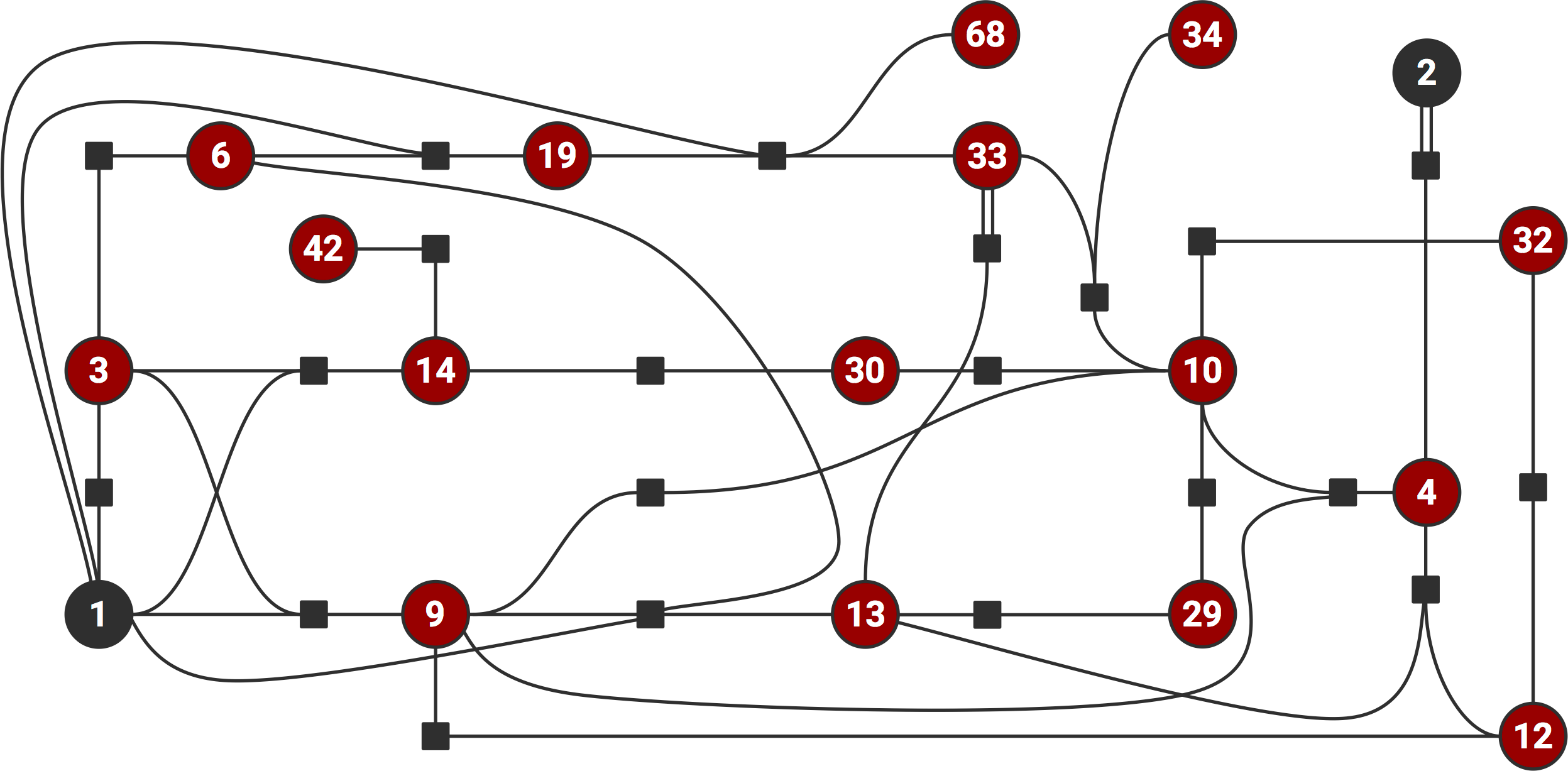}
\caption{
Sparse variant of the exemplary reaction network CRN-X consisting of 18 vertices (numbered circles) and 21 edges.
The center of every edge is represented by a square, which may be interpreted as transition state.
Two sides of each square are linked with either one or two lines, which are, in turn, linked with a single vertex each.
The left-hand-side and right-hand-side vertices of a reaction are never connected with the same side of a square.
Vertices corresponding to reactants are black-colored, whereas all other vertices are red-colored.
}
\label{fig:full}
\end{figure}

	Given a concentration error tolerance of $\varepsilon_\mathbf{y} = 0.01$~mol~L$^{-1}$, we find that a DFA-based network reduction with $G_\text{min} = 1.0\times10^{-3}$~mol~L$^{-1}$ is reliable with respect to both $\delta_{pq}(t_\text{max})$ and $\Delta_{pq}$.
	Here, we chose the minimum confidence level of $\gamma = 0$ for the sake of clarity:
	The smaller $\gamma$, the larger the possible degree of network reduction, which leads to more comprehensible reaction mechanisms.
	In an actual setup, we would recommend to choose a larger confidence level.
	Note that we chose $G_\text{min}$ to be one magnitude smaller than $\varepsilon_\mathbf{y}$.
	The reason is that $G_\text{min}$ is assessed on a species-by-species basis, whereas $\varepsilon_\mathbf{y}$ is compared against the quantities $\delta_{pq}(t_\text{max})$ and $\Delta_{pq}$, which measure the concentration error summed over all species.
	One may resolve this situation by dividing $G_\text{min}$ by $N$, which is rather conservative (i.e., it would lower the possible degree of network reduction) but also more reliable (i.e., the resulting concentration error would be smaller).
	Fig.~\ref{fig:full} shows the sparse variant of CRN-X, which only consists of 18 vertices and 21 edges, corresponding to about 20\,\% of the network elements contained in the original network.
	Note that the number of kinetically relevant species identified by our DFA analysis varies between 17 and 21 if the 25 solutions are considered individually.
	Hence, the explicit consideration of \textbf{k}-uncertainty has a direct effect on the degree of network reduction in this case.
	Note that the possible degree of reduction becomes larger (on average) for an increasing number of vertices and edges.
	
\begin{figure} [!ht]
\centering
\includegraphics[width=\textwidth]{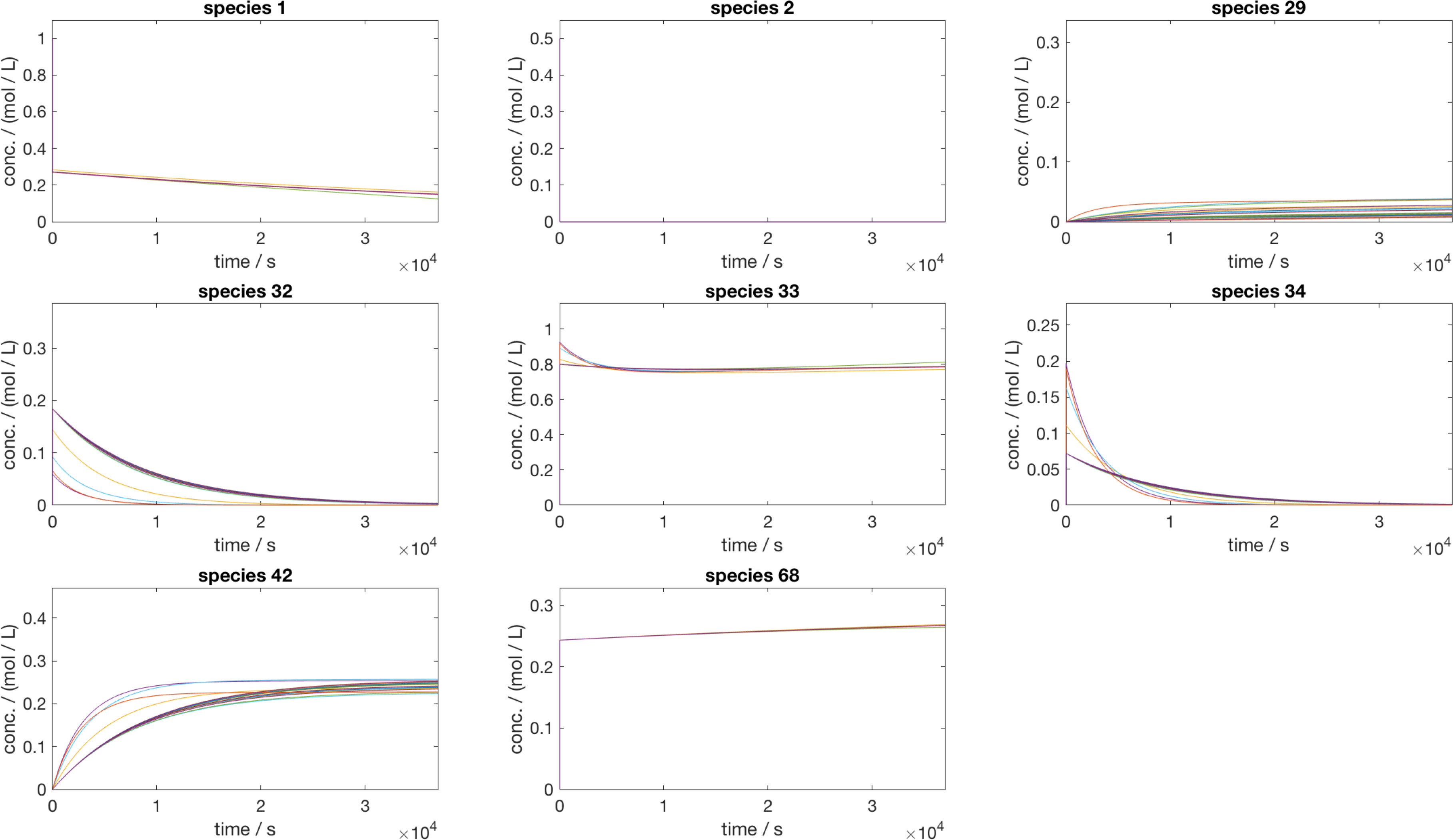}
\caption{
Concentration--time plots for the dominant species of the exemplary reaction network CRN-X \textit{after} refinement of 9 pairs of rate constants.
An ensemble of 25 distinct kinetic models derived from CRN-X has been considered.
}
\label{fig:post_msa}
\end{figure}

	The combination of a large \textbf{y}-uncertainty and a still quite entangled network renders it difficult to suggest a specific reaction mechanism.
	To resolve this issue, we conducted a global sensitivity analysis based on our extended Morris screening approach.
	For $C = 5$ Morris samples, our analysis suggests 9 out of 21 reaction pairs to be critical, i.e., they yield values for $\delta_{pq}(t_\text{max})$ and/or $\Delta_{pq}$ exceeding $\varepsilon_\mathbf{y}$ with respect to the quantile specified by $\gamma$.
	Assessing the 5 Morris samples one by one, the number of critical reactions varies between 7 and 13.
	Here, we simulated the refinement of rate constants by taking the averages of the absolute free energies in question (for both vertices and edges) over all 25 samples, which resulted in rate constants with zero variance for the 9 critical reaction pairs.
	The corresponding concentration--time plots are shown in Fig.~\ref{fig:post_msa}.
	The interpretability of the solutions increased significantly.
	Species \textbf{33} is indeed the main product with a final concentration of about 0.8~mol~L$^{-1}$.
	Furthermore, species \textbf{42} and \textbf{68} are relevant side products with final concentrations of 0.2--0.3~mol~L$^{-1}$, whereas the final concentrations of species \textbf{29}, \textbf{32}, and \textbf{34} are less significant.
	
\begin{figure} [!ht]
\centering
\includegraphics[width=\textwidth]{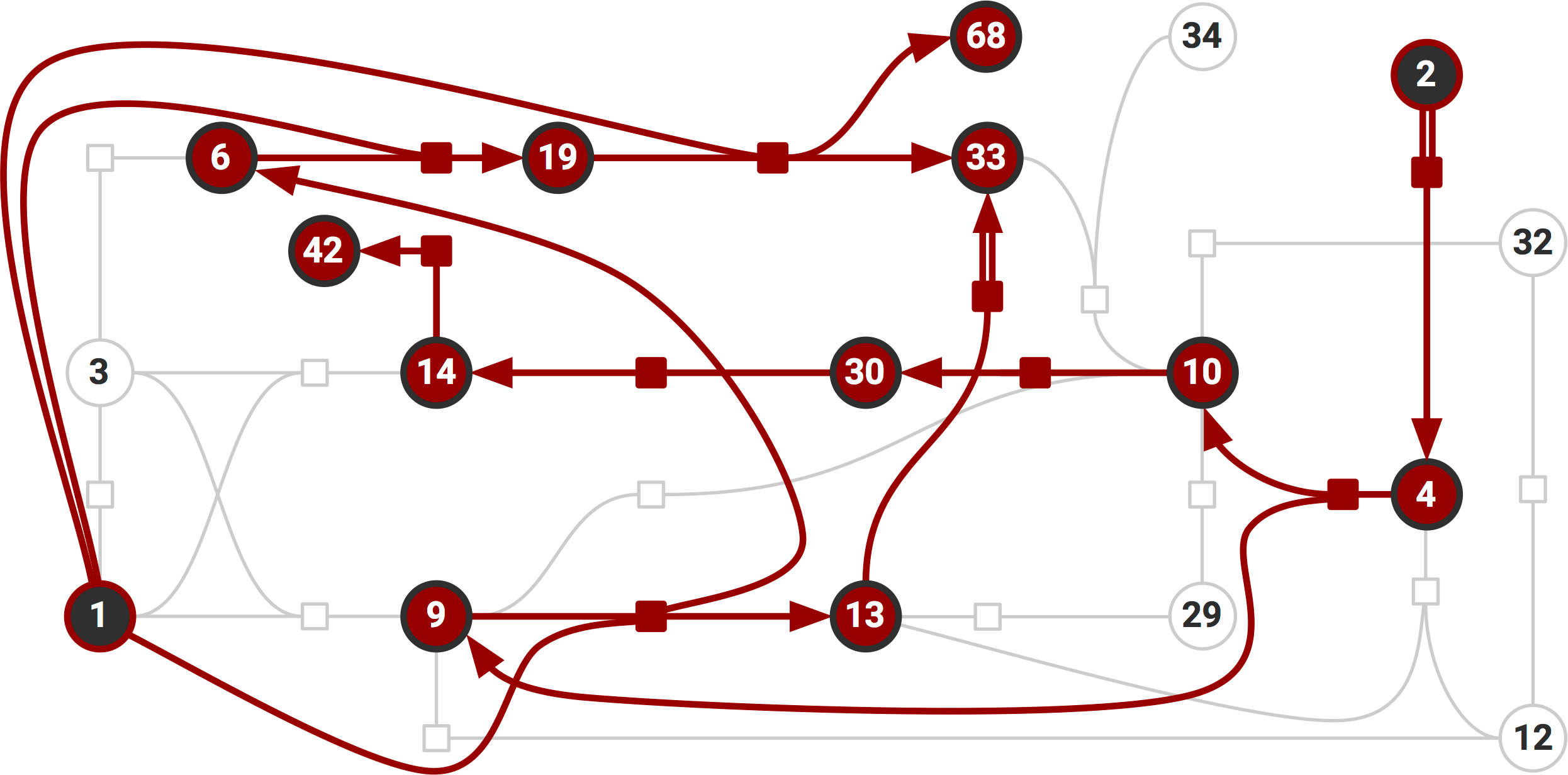}
\caption{
Sparse variant of CRN-X illustrating the dominant reaction paths (red-colored edges with arrow heads).
All species that are part of dominant reaction paths are represented by red-colored vertices except for reactant vertices which are black-colored.
}
\label{fig:dominant}
\end{figure}

	When analyzing the edge flux \textit{quanta}, i.e., the edge fluxes for a given time interval $t_u - t_{u-1}$, we can reconstruct the dominant reaction pathways leading to the main and side products (Fig.~\ref{fig:dominant}).
	In the very beginning of the reaction, \textbf{2} dimerizes immediately and completely to \textbf{4}, which, in turn, immediately and completely dissociates to \textbf{9} and \textbf{10}.
	The formation of \textbf{9} enables its reaction with \textbf{1} to form \textbf{6} and \textbf{13}, the latter of which reacts quickly to \textbf{33}, the main product.
	Of the three reaction channels that lead to the formation of \textbf{33}, this channel is the most important one.
	The formation of \textbf{6} via \textbf{1} and \textbf{9} activates its reaction with \textbf{1} to \textbf{19}, the latter two of which react further to \textbf{33} and \textbf{68}, one of the two dominant side products.
	On a longer time scale, \textbf{10} reacts to \textbf{42}\,---\,the other dominant side product\,---\,via \textbf{30} and \textbf{14}.
	In summary, the dominant reaction pathways read:
	
\textbf{2} + \textbf{2} $\rightarrow$ \textbf{4} $\rightarrow$ \textbf{9} + \textbf{10},

\textbf{1} + \textbf{9} $\rightarrow$ \textbf{6} + \textbf{13},

\textbf{13} $\rightarrow$ \textbf{33} + \textbf{33},

\textbf{1} + \textbf{6} $\rightarrow$ \textbf{19},

\textbf{1} + \textbf{19} $\rightarrow$ \textbf{33} + \textbf{68},

\textbf{10} $\rightarrow$ \textbf{30} $\rightarrow$ \textbf{14} $\rightarrow$ \textbf{42}.

	In a practical setup, one would conduct the kinetic analysis presented above not only at the end but rather repeatedly during the exploration as many of the intermediates and transition states may be kinetically irrelevant.
	The number of such redundant states potentially grows superlinearly with the size of the network since every vertex that can only be reached via irrelevant channels will be irrelevant as well.
	Combined with the fact that the final network size is unknown a priori, the coupling of kinetic modeling with network exploration is generally crucial.
	
	Here, we applied the algorithm presented in Section~\ref{sec:exploration} to CRN-X, but performed network reduction and sensitivity analysis only at the end of the exploration.
	As already mentioned, potentially relevant edges may reveal small fluxes in early stages of the exploration, which renders any flux-based network reduction critical.
	Instead of choosing a rate threshold as a completeness criterion, we stopped the exploration when the solution of the current network did not differ from the solution of the full network by more than $\varepsilon_\mathbf{y}$ as measured by $\delta_{pq}(t_\text{max})$ and $\Delta_{pq}$.
	Hence, we needed to switch off the sensitivity analysis during exploration as it would have led to incomparable solutions.
	
	The \textsc{KiNetX}-guided exploration of CRN-X required 17 steps, leading to 63 vertices and 68 edges, which corresponds to an effective network reduction of about 40\,\%, which is significantly below the 80\,\% we obtained with our DFA algorithm.
	However, choosing a conservative exploration strategy that does not make a priori assumptions about the underlying mechanism, one cannot bypass a certain degree of redundancy.

\subsection{Relevance of Uncertainty Propagation}
	To assess the importance of explicitly considering \textbf{k}-uncertainty in  kinetic studies of room-temperature solution chemistry, we need to examine a multitude of chemical reaction networks for which correlated uncertainty information is available.
	Currently, the data situation is still too poor to resort to reaction networks generated with chemistry-based exploration codes.
	For this reason, we randomly created 100 reaction networks with \textsf{AutoNetGen}.
	Each network contains precisely 100 edges and is based on two reactants with the same initial concentrations and masses introduced in Section~\ref{sec:xmpl}.
	All remaining input values are listed in Table~\ref{tab:input}.
	The average number of vertices in these networks explored without kinetic guidance is 45 ($\pm6$).
	The ensemble-averaged activation free-energy uncertainty amounts to 1.0~kcal~mol$^{-1}$ with a standard deviation of 0.4~kcal~mol$^{-1}$. 
	This uncertainty range is rather small compared to what we estimated for small-molecule organic chemistry with the LC$^\ast$-PBE0 density functional ensemble.\cite{proppe2016}
	We conducted the kinetic analysis in two different ways.
	In the first case, we only considered the nominal kinetic model based on $\mathbf{k}_0$.
	We refer to this case as CRN-0.
	In the second case named CRN-B, we considered the nominal model plus 5 additional sampled models.
	This way, we can estimate the importance of incorporating uncertainty propagation into kinetic modeling.
	The results of both analyses are summarized in Table~\ref{tab:results}.
	
\begin{table} [!ht]
\center
\caption{
Mean values and standard devations of properties obtained as a result of \textsc{KiNetX}-based analyses on the CRN-0 and CRN-B cases.
}
\vspace{0.2cm}
\label{tab:results}
\begin{tabular}{lrrrr}
\hline\hline
Property, $\zeta$ & $\mu(\zeta_\text{CRN-0})$ & $\mu(\zeta_\text{CRN-B})$ & $\sigma(\zeta_\text{CRN-0})$ & $\sigma(\zeta_\text{CRN-B})$ \\
\hline\hline
\#exploration steps &12&14&7&7\\
\#critical reactions &5&10&4&7\\
$L_\text{expl}$ &54&60&26&27\\
$N_\text{expl}$ &29&31&11 & 11\\
$L_\text{red}$ &30&37&18&21\\
$N_\text{red}$ &17&20&8&8\\
$\delta_{B,\text{expl}}(t_\text{max})$ / mol~L$^{-1}$ &---&0.14&---&0.13\\
$\Delta_{B,\text{expl}}$ / mol~L$^{-1}$&---&0.15&---&0.13\\
$\delta_{B,\text{refined}}(t_\text{max})$ / mol~L$^{-1}$&---&$<$0.01&---&0.01\\
$\Delta_{B,\text{refined}}$ / mol~L$^{-1}$&---&$<$0.01&---&0.01\\
\hline\hline
\end{tabular}
\end{table}
	
	In case of exploration guidance by \textsc{KiNetX}, the average number of vertices and edges decreases to 31 and 60 for CRN-B, respectively, which corresponds to a reduction of network elements of about 35\,\%.
	The number of vertices and edges in the case of CRN-0 is slightly smaller (29 and 54, respectively), which is to be expected since the consideration of several instead of a single solution naturally increases the number of potentially relevant reaction channels.
	This also indicates why two additional exploration steps were necessary on average in the CRN-B case.
	Additionally, we recorded the maximum formation rate initiating the last exploration for each network.
	The minimum over all networks amounts to $1.0\times10^{-17}$~mol~L$^{-1}$~s$^{-1}$ in the CRN-B case.
	If we choose the maximum possible time interval (the time of reaction, $t_\text{max}$), we obtain a flux of $3.7\times10^{-13}$~mol~L$^{-1}$, which is very small compared to $\varepsilon_\mathbf{y} = 0.01$~mol~L$^{-1}$.
	This finding confirms the limitations of the rate-based algorithm discussed in Section~\ref{sec:exploration}:
	There are cases in which the rate threshold would need to be chosen so small that nothing is gained by coupling a kinetic analysis to network exploration.
	However, one does not know a priori when this situation occurs.
	
	DFA-based network reduction with a flux threshold of $G_\text{min} = 1.0\times10^{-5}$~mol~L$^{-1}$ was successful for 83 networks in the CRN-B case, whereas it was successful for only 78 networks in the CRN-0 case.
	Similar to our argument of the last paragraph, the consideration of several solutions increases the likelihood to identify kinetically relevant network elements.
	Therefore, we also find that the resulting number of vertices and edges is larger in the CRN-B case after network reduction (including LBA-based reduction).
	Note that the reduction percentage of approximately 40\,\% (75\,\% compared to the networks explored without kinetic guidance) is likely to become larger for an increasing size of the original network.
	To test this hypothesis, we chose the same 100 random seeds employed for the generation of the 100-edge networks but increased the number of edges to 500.
	Neglecting kinetic guidance during exploration and considering only DFA-based reduction, we find an average increase in the reduction of network elements from about 75\,\% to 90\,\%, which confirms our hypothesis.
	
	We find that, on average, 10 reactions per network are critical in the CRN-B case, corresponding to 28\,\% of the reactions.
	Analogously to the argument provided with regards to the possible degree of network reduction, we expect the number of critical reactions identified for an ensemble of kinetic models to be larger than for a single model.
	Indeed, only 5 reactions per network (on average) were found to be critical in the CRN-0 case.
	After global sensitivity analysis and refinement of activation free energies, we find an average decrease of $\max\big\{\delta_B(t_\text{max}),\Delta_B\big\}$ from 0.15~mol~L$^{-1}$ to ${<0.01}$~mol~L$^{-1}$ for the CRN-B case, which highlights the success of our extended Morris screening approach.
	The refinement of activation free energies was mimicked by taking the mean value of absolute free energies (for both vertices and edges) over all $B + 1 = 6$ samples for the critical reactions.
	This way, the resulting activation free energies of (at least) the critical reactions are identical for all kinetic models.
	It is important to manipulate absolute instead of relative free energies.
	Otherwise, one may induce unphysical scenarios by violating the necessary condition of microscopic reversibility.
	
	Finally, we examined the reliability of the solutions obtained for CRN-0 and CRN-B after a series of kinetics-guided exploration, network reduction, and free-energy refinement.
	For this purpose, we generated an ``exact'' solution obtained from taking the mean value of absolute free energies over \textit{all} reactions of the full network explored without kinetic guidance.
	The average value of $\max\big\{\delta_{pq}(t_\text{max}),\Delta_{pq}\big\}$ comparing the CRN-0 solutions against the ``exact'' solutions amounts to $15.7\times10^{-3}$~mol~L$^{-1}$, whereas it amounts to only $1.3\times10^{-3}$~mol~L$^{-1}$ when comparing the ensemble-averaged CRN-B solutions against the ``exact'' solutions.


\section{Conclusions and Outlook}

	In this paper, we have demonstrated the strong capability of advanced kinetic modeling techniques for the deduction of product distributions, reaction mechanisms, and possibly other properties of chemical systems from reaction data equipped with correlated uncertainty information.
	Our approach is designed (but not limited) to be fed with raw data from quantum chemical calculations as we aim to develop a flexible kinetic modeling framework rooted in the first principles of quantum mechanics.
	For this purpose, we developed the meta-algorithm \textsc{KiNetX}, which carries out kinetic analyses of complex chemical reaction networks in a fully automated manner, including guidance for network exploration, hierarchical network reduction, and global sensitivity analysis.
	The key feature of \textsc{KiNetX} is that the correlated uncertainties of the model parameters (activation free energies or rate constants) are rigorously propagated through all steps of the kinetic modeling workflow.
	
	We demonstrated the entire workflow of \textsc{KiNetX} at a reaction network generated with \textsf{AutoNetGen}, an algorithm which constructs artificial reaction networks by encoding chemical logic into their underlying graph structure.
	Our results show that \textsc{KiNetX} can systematically interpret noisy concentration data to guide the exploration of reaction space and  identify regions in a network that require more accurate free-energy data, without the need to carry out high-accuracy quantum chemical calculations for all species considered.
	Furthermore, we showed that the dominant reaction pathways can be reliably deduced as a result of these efforts.
	To study the reliability increase by incorporating uncertainty quantification into the kinetic modeling framework, we were faced with the challenge to generate a multitude of reaction networks for covering a wide chemical spectrum, which is very time-consuming regarding the number of quantum chemical calculations required for this purpose.
	With the development of \textsf{AutoNetGen}, we were able to examine a large number of distinct chemistry-like scenarios in short time.
	Here, we considered 100 networks consisting of 100 edges and 45 ($\pm6$) vertices and distinguished between two cases.
	In the first case, we only considered a single kinetic model per network.
	In the second case, we considered an additional number of 5 kinetic models per network.
	Our results suggest, despite the small number of samples considered in the second case, that the rigorous propagation of uncertainty through all steps of a kinetic modeling study can significantly increase the reliability of conclusions; here, by a factor of 10.
	
	While the findings are appealing, we understand that the use of chemistry-mimicking reaction networks requires a careful analysis to ensure that important network properties met in actual chemical scenarios are captured.
	Otherwise, it may be ambiguous to generalize our conclusions drawn from these artificial cases.
	At the same time, it is difficult to draw general conclusions about certain network properties\,---\,such as the percentage of critical (highly noise-inducing) reactions, the potential degree of network reduction, or the number of network layers required to correctly account for all kinetically relevant reaction steps\,---\,as they may be strongly dependent on the underlying graph structure, the distribution of activation free energies / rate constants, as well as their correlated uncertainties.
	It is not known to us that the literature on solution chemistry (or chemistry in general) offers enough data in this direction to reliably evaluate this dependency.
	Recent results from our group suggest that free-energy uncertainty may induce almost arbitrary magnitudes of concentration noise \cite{proppe2016}.
	We developed AutoNetGen to offer a more general (i.e., statistics-based) perspective on this important issue despite a poor data situation.
	There are a few arguments in favor of our artificial, chemistry-mimicking reaction networks.
	First, AutoNetGen is, in principle, able to generate every network graph that corresponds to a specific chemical reaction characterized by elementary processes with a molecularity smaller than three.
	Second, we coordinated the range of activation free energies (0--100~kJ~mol$^{-1}$ with respect to the higher-energy intermediate) with the reaction time, $t_\text{max}$, which we set to the half life of a unimolecular rate constant corresponding to a barrier height of 100~kJ~mol$^{-1}$.
	This way, we avoid that the network reduction procedure merely deletes reactions because of activation barriers that are too high in energy. 
	Note that in actual chemical systems, several activation barriers may be much higher in energy and, hence, we expect the degree of network reduction to be rather small compared to actual chemical scenarios.
	Third, the activation free-energy uncertainties studied here amount to (1.0 $\pm$ 0.4)~kcal~mol$^{-1}$, which is rather small compared to what we found for actual activation barriers obtained from DFT calculations.\cite{proppe2016}
	Fourth, our analysis consistently suggests that the explicit consideration of ensembles of kinetic models improves the reliability of conclusions for a diverse range of networks.
	In total, there is some indication that our chemistry-mimicking reaction networks provide a trend for what is to be expected if rigorous uncertainty propagation is incorporated in the kinetic analysis of actual chemical systems.
	Certainly, the application of \textsc{KiNetX} to real-world examples (not only by us but by the entire community) is our long-term goal, but it is outside the scope of this paper. 
	In future work, we will couple \textsc{KiNetX} to our automated network exploration program \textsc{Chemoton}\cite{simm2017a} to assess the performance of \textsc{KiNetX} with respect to relevant examples such as the very challenging autocatalytic formose reaction.
	Both projects are part of our developments for a new kind of computational quantum chemistry (SCINE).\cite{scine}


\section*{Acknowledgments}
The authors gratefully acknowledge financial support from the Swiss National Science Foundation (project no.\ 200020\_169120).


\section*{Appendix}
\subsection*{Details on \textsf{AutoNetGen}}
	We require \textsf{AutoNetGen} to yield a fully connected network representing unimolecular reactions (isomerization and dissociation) as well as bimolecular reactions (dissociation and substitution).
	\textsf{AutoNetGen} requires the specification of reactants including their masses.
	The following steps are carried out by \textsf{AutoNetGen} for the construction of artificial reaction networks:
	\begin{enumerate}
	\item For the construction of the $(x+1)$-th network layer, we first define all potentially reactive intermediate states. At that stage, we register $N = N_0 + \cdots + N_x$ vertices. 
	Since we already explored all possible reactions between the first $N - N_x$ species, we will only consider the following potentially reactive intermediate states: 
	$N_x$ unimolecular intermediate states (leading to reactions of type A~$\rightarrow$~P) defined by the $N_x$ species of the $x$-th network layer, $N_x$ homobimolecular intermediate states (type 2A~$\rightarrow$~P), and $N_x(N_x -1)/2 + N_x(N-N_x)$ heterobimolecular intermediate states (type A + B~$\rightarrow$~P).
	The first $N_x(N_x -1)/2$ heterobimolecular states represent all possible, nonredundant combinations of the $N_x$ new species among each other, whereas the latter $N_x(N-N_x)$ represent all possible combinations between the $N_x$ new species and the $N-N_x$ old species.
	\item For each of the potentially reactive intermediate states, we draw from an exponential distribution with mean $\mu_\text{exp}$.
	The user can specify two different values for $\mu_\text{exp}$, one for unimolecular reactions, $\mu_\text{exp,uni}$, and one for bimolecular reactions, $\mu_\text{exp,bi}$.
	The floor value of the sampled value equals the number of reaction channels to be explored from the current reactive intermediate state.
	The specific number of reaction channels determines how many distinct reactive transformations of the corresponding intermediate state will occur and, therefore, how many new vertices will be formed or vertices of an earlier layer will be reconnected.
	The user can control the tendency to generate new vertices over linking to vertices of earlier layers via the parameter $p(\mathcal{V}_\text{new})$ ranging from zero to one.
	If $p(\mathcal{V}_\text{new}) = 0$, AutoNetGen will never generate a new vertex, whereas it will never link to a vertex of an earlier layer if $p(\mathcal{V}_\text{new}) = 1$.
	\textsf{AutoNetGen} ensures, of course, that the mass on both sides of a reaction is always identical.
	
	\item When the exploration stops (e.g., when a user-defined number of edges is reached), $2L$ activation free energies will be calculated:
	The absolute free energy of the product (or right-hand) side of an edge (comprises either one or two vertices) is sampled from a normal distribution with mean $\mu( A^- - A^+)$ and standard deviation $\sigma( A^- - A^+)$, which is added to the absolute free energy of the reactant (or left-hand) side of that edge.
	The absolute free energy of an edge is sampled from a uniform distribution with bounds $\min (A^\ddagger - A^{+/-})$ and $\max (A^\ddagger - A^{+/-})$, which is added to the higher-energy side of an edge.
	The differences between edge and vertex energies are the activation free energies of the underlying reaction network.
	
	\item We introduce free-energy uncertainties in two ways. First, we sample vertex free energies from their nominal values and the covariance matrix $\boldsymbol{\Sigma}_\mathbf{A}$ as described in the next section.
	Second, for a given reaction and a given sample, we add the mean value of the free-energy changes in the two connected intermediate states (compared to their nominal values) to the free energy of the corresponding edge and add another contribution to it which is sampled from a normal distribution with zero mean and standard deviation $2\langle \sigma_A \rangle$.
	In case the resulting activation free energy becomes negative, its absolute value will be chosen.
	\end{enumerate}
	
	Subsequently, \textsf{AutoNetGen} transforms all activation free energies to rate constants based on the Eyring equation,\cite{eyring1935}
\begin{equation}
\label{eq:eyring}
k_l^{+/-} = \frac{k_\text{B}T}{h} \exp \Bigg(-\frac{A_l^\ddagger - A_l^{+/-}}{RT} \Bigg) \ ,
\end{equation}
where $h$, $k_\text{B}$, $R$, and $T$ are Planck's constant, Boltzmann's constant, the gas constant, and a user-defined constant temperature, respectively.

\subsection*{Random Construction of Covariance Matrices}
	Covariance matrices are symmetric, positive-semidefinite square matrices by definition, i.e., their eigenvalues are strictly nonnegative.
	We outline a simple recipe to construct a random covariance matrix $\boldsymbol{\Sigma}_\mathbf{A}$, which fulfills the condition that its largest eigenvalue equals $\sigma_{A_\text{max}}^2$.
	Defining $\sigma_{A_\text{max}}^2$ to be the largest eigenvalue ensures that all activation free-energy uncertainties are bound by $\sigma_{A_\text{max}}$.
	
	\begin{enumerate}
	\item Generate a random $(N \times N)$-dimensional matrix $\mathbf{P}$ with elements that are uniformly sampled from $[-0.5,+0.5]$. $N$ is the number of vertices in the network.
	\item The multiplication of $\mathbf{P}$ with its transpose leads to a $(N \times N)$-dimensional matrix $\mathbf{Q} = \mathbf{P} \mathbf{P}^\top$ that already is a covariance matrix\cite{horn1990} with eigenvalues $\lbrace E_{ii} \rbrace$. 
	\item The covariance matrix $\boldsymbol{\Sigma}_\mathbf{A}$ results from a rescaling of $\mathbf{Q}$,
	\begin{equation}
	\boldsymbol{\Sigma}_\mathbf{A} = \mathbf{Q} \frac{\langle \sigma_A \rangle^2}{\max \lbrace E_{ii} \rbrace} \ , \nonumber
	\end{equation}
	which implies that the largest eigenvalue of $\boldsymbol{\Sigma}_\mathbf{A}$ equals $\langle \sigma_A \rangle^2$. 
	\end{enumerate}
	
	Note that, in a practical setup, we expect the user to provide $B+1$ \textbf{k}-vectors derived from an ensemble of quantum chemical models, instead of sampling the corresponding activation free energies from a random (and, hence, problem-unrelated) covariance matrix.
	However, as it is current practice to generate a single set of activation free energies instead of an ensemble of them, we encourage users to employ these random covariance matrices to develop an intuition for the \textit{potential} effect of free-energy uncertainty on the kinetics of complex chemical systems.
	
\clearpage
\subsection*{Control Parameters for \textsf{AutoNetGen} and \textsc{KiNetX}}	

Table \ref{inputpara} provides an overview of all parameters that are required for
\textsf{AutoNetGen} and \textsc{KiNetX} on input and contains the values chosen for the three cases discussed in this work.
	
\begin{table} [!ht]
\center
\caption{
Input parameters for \textsf{AutoNetGen} and \textsc{KiNetX} and their values chosen for this study.
\label{inputpara}
}
\vspace{0.2cm}
\label{tab:input}
\begin{tabular}{lrrr}
\hline\hline
Input parameter & CRN-X & CRN-0 & CRN-B \\
\hline\hline
\textsf{AutoNetGen} \\
\hline
$B+1$ &25&1&6\\
$T$ / K &298.15&298.15&298.15\\
$\mu( A^- - A^+)$ / (kJ~mol$^{-1}$) &$-$25&$-$25&$-$25\\
$\sigma( A^- - A^+)$ / (kJ~mol$^{-1}$) &50&50&50\\
$\min (A^\ddagger - A^{+/-})$ / (kJ~mol$^{-1}$) &0&0&0\\
$\max (A^\ddagger - A^{+/-})$ / (kJ~mol$^{-1}$) &100&100&100\\
$\langle \sigma_A \rangle$ / (kJ~mol$^{-1}$) &2.09&2.09&2.09\\
$L_\text{max}$ &125&100&100\\
$\mu_\text{exp}(N_\text{uni})$ &5&5&5\\
$\mu_\text{exp}(N_\text{bi})$ &2&2&2\\
$p(\mathcal{V}_\text{new})$ &0.5&0.1&0.1\\
\hline\hline
\textsc{KiNetX} \\
\hline
$t_\text{max}$ / s &36954&36954&36954\\
$\varepsilon_\mathbf{y}$ / (mol~L$^{-1}$) &0.01&0.01&0.01\\
$G_\text{min}$ / (mol~L$^{-1}$) &$1.0\times10^{-3}$&$1.0\times10^{-5}$&$1.0\times10^{-5}$\\
$U$ &1000&1000&1000\\
$\gamma$ &0&1&1\\
$C$ &5&1&5\\
\hline\hline
\end{tabular}
\end{table}
	

\providecommand{\refin}[1]{\\ \textbf{Referenced in:} #1}



\begin{thebibliography}{10}

\bibitem{broadbelt1994}
Broadbelt,~L.~J.;\ \ Stark,~S.~M.;\ \ Klein,~M.~T.  Computer {{Generated
  Pyrolysis Modeling}}: {{On}}-the-{{Fly Generation}} of {{Species}},
  {{Reactions}}, and {{Rates}},  \textit{Ind. Eng. Chem. Res.} \textbf{1994,}
  \textsl{33,} 790--799.

\bibitem{kayala2012}
Kayala,~M.~A.;\ \ Baldi,~P.  {{ReactionPredictor}}: {{Prediction}} of {{Complex
  Chemical Reactions}} at the {{Mechanistic Level Using Machine Learning}},
  \textit{J. Chem. Inf. Model.} \textbf{2012,} \textsl{52,} 2526--2540.

\bibitem{maeda2013}
Maeda,~S.;\ \ Ohno,~K.;\ \ Morokuma,~K.  Systematic Exploration of the
  Mechanism of Chemical Reactions: The Global Reaction Route Mapping ({{GRRM}})
  Strategy Using the {{ADDF}} and {{AFIR}} Methods,  \textit{Phys. Chem. Chem.
  Phys.} \textbf{2013,} \textsl{15,} 3683--3701.

\bibitem{zimmerman2013}
Zimmerman,~P.~M.  Automated Discovery of Chemically Reasonable Elementary
  Reaction Steps,  \textit{J. Comput. Chem.} \textbf{2013,} \textsl{34,}
  1385--1392.

\bibitem{rappoport2014}
Rappoport,~D.;\ \ Galvin,~C.~J.;\ \ Zubarev,~D.~Y.;\ \ {Aspuru-Guzik},~A.
  Complex {{Chemical Reaction Networks}} from {{Heuristics}}-{{Aided Quantum
  Chemistry}},  \textit{J. Chem. Theory Comput.} \textbf{2014,} \textsl{10,}
  897--907.

\bibitem{wang2014}
Wang,~L.-P.;\ \ Titov,~A.;\ \ McGibbon,~R.;\ \ Liu,~F.;\ \ Pande,~V.~S.;\ \
  Mart\'inez,~T.~J.  Discovering Chemistry with an Ab Initio Nanoreactor,
  \textit{Nat. Chem.} \textbf{2014,} \textsl{6,} 1044--1048.

\bibitem{bergeler2015}
Bergeler,~M.;\ \ Simm,~G.~N.;\ \ Proppe,~J.;\ \ Reiher,~M.  Heuristics-{{Guided
  Exploration}} of {{Reaction Mechanisms}},  \textit{J. Chem. Theory Comput.}
  \textbf{2015,} \textsl{11,} 5712--5722.

\bibitem{dontgen2015}
D\"ontgen,~M.;\ \ {Przybylski-Freund},~M.-D.;\ \ Kr\"oger,~L.~C.;\ \
  Kopp,~W.~A.;\ \ Ismail,~A.~E.;\ \ Leonhard,~K.  Automated {{Discovery}} of
  {{Reaction Pathways}}, {{Rate Constants}}, and {{Transition States Using
  Reactive Molecular Dynamics Simulations}},  \textit{J. Chem. Theory Comput.}
  \textbf{2015,} \textsl{11,} 2517--2524.

\bibitem{habershon2015}
Habershon,~S.  Sampling Reactive Pathways with Random Walks in Chemical Space:
  {{Applications}} to Molecular Dissociation and Catalysis,  \textit{J. Chem.
  Phys.} \textbf{2015,} \textsl{143,} 094106.

\bibitem{suleimanov2015}
Suleimanov,~Y.~V.;\ \ Green,~W.~H.  Automated {{Discovery}} of {{Elementary
  Chemical Reaction Steps Using Freezing String}} and {{Berny Optimization
  Methods}},  \textit{J. Chem. Theory Comput.} \textbf{2015,} \textsl{11,}
  4248--4259.

\bibitem{zimmerman2015}
Zimmerman,~P.~M.  Navigating Molecular Space for Reaction Mechanisms: An
  Efficient, Automated Procedure,  \textit{Mol. Simul.} \textbf{2015,}
  \textsl{41,} 43--54.

\bibitem{dewyer2017}
Dewyer,~A.~L.;\ \ Zimmerman,~P.~M.  Finding Reaction Mechanisms, Intuitive or
  Otherwise,  \textit{Org. Biomol. Chem.} \textbf{2017,} \textsl{15,} 501--504.

\bibitem{gao2016}
Gao,~C.~W.;\ \ Allen,~J.~W.;\ \ Green,~W.~H.;\ \ West,~R.~H.  Reaction
  {{Mechanism Generator}}: {{Automatic}} Construction of Chemical Kinetic
  Mechanisms,  \textit{Comput. Phys. Commun.} \textbf{2016,} \textsl{203,}
  212--225.

\bibitem{habershon2016}
Habershon,~S.  Automated {{Prediction}} of {{Catalytic Mechanism}} and {{Rate
  Law Using Graph}}-{{Based Reaction Path Sampling}},  \textit{J. Chem. Theory
  Comput.} \textbf{2016,} \textsl{12,} 1786--1798.

\bibitem{wang2016}
Wang,~L.-P.;\ \ McGibbon,~R.~T.;\ \ Pande,~V.~S.;\ \ Martinez,~T.~J.  Automated
  {{Discovery}} and {{Refinement}} of {{Reactive Molecular Dynamics Pathways}},
   \textit{J. Chem. Theory Comput.} \textbf{2016,} \textsl{12,} 638--649.

\bibitem{simm2017a}
Simm,~G.~N.;\ \ Reiher,~M.  Context-{{Driven Exploration}} of {{Complex
  Chemical Reaction Networks}},  \textit{J. Chem. Theory Comput.}
  \textbf{2017,} \textsl{13,} 6108--6119.

\bibitem{dontgen2018}
D\"ontgen,~M.;\ \ Schmalz,~F.;\ \ Kopp,~W.~A.;\ \ Kr\"oger,~L.~C.;\ \
  Leonhard,~K.  Automated {{Chemical Kinetic Modeling}} via {{Hybrid Reactive
  Molecular Dynamics}} and {{Quantum Chemistry Simulations}},  \textit{J. Chem.
  Inf. Model.} \textbf{2018,} \textsl{58,} 1343--1355.

\bibitem{dewyer2018}
Dewyer,~A.~L.;\ \ Arg\"uelles,~A.~J.;\ \ Zimmerman,~P.~M.  Methods for
  Exploring Reaction Space in Molecular Systems,  \textit{WIREs Comput. Mol.
  Sci.} \textbf{2018,} \textsl{8,} e1354.

\bibitem{frederiksen2004}
Frederiksen,~S.~L.;\ \ Jacobsen,~K.~W.;\ \ Brown,~K.~S.;\ \ Sethna,~J.~P.
  Bayesian {{Ensemble Approach}} to {{Error Estimation}} of {{Interatomic
  Potentials}},  \textit{Phys. Rev. Lett.} \textbf{2004,} \textsl{93,} 165501.

\bibitem{mortensen2005}
Mortensen,~J.~J.;\ \ Kaasbjerg,~K.;\ \ Frederiksen,~S.~L.;\ \
  N\o{}rskov,~J.~K.;\ \ Sethna,~J.~P.;\ \ Jacobsen,~K.~W.  Bayesian {{Error
  Estimation}} in {{Density}}-{{Functional Theory}},  \textit{Phys. Rev. Lett.}
  \textbf{2005,} \textsl{95,} 216401.

\bibitem{petzold2012}
Petzold,~V.;\ \ Bligaard,~T.;\ \ Jacobsen,~K.~W.  Construction of {{New
  Electronic Density Functionals}} with {{Error Estimation Through Fitting}},
  \textit{Top. Catal.} \textbf{2012,} \textsl{55,} 402--417.

\bibitem{wellendorff2012}
Wellendorff,~J.;\ \ Lundgaard,~K.~T.;\ \ M\o{}gelh\o{}j,~A.;\ \ Petzold,~V.;\ \
  Landis,~D.~D.;\ \ N\o{}rskov,~J.~K.;\ \ Bligaard,~T.;\ \ Jacobsen,~K.~W.
  Density Functionals for Surface Science: {{Exchange}}-Correlation Model
  Development with {{Bayesian}} Error Estimation,  \textit{Phys. Rev. B}
  \textbf{2012,} \textsl{85,} 235149.

\bibitem{medford2014}
Medford,~A.~J.;\ \ Wellendorff,~J.;\ \ Vojvodic,~A.;\ \ Studt,~F.;\ \
  {Abild-Pedersen},~F.;\ \ Jacobsen,~K.~W.;\ \ Bligaard,~T.;\ \
  N\o{}rskov,~J.~K.  Assessing the Reliability of Calculated Catalytic Ammonia
  Synthesis Rates,  \textit{Science} \textbf{2014,} \textsl{345,} 197--200.

\bibitem{pandey2015}
Pandey,~M.;\ \ Jacobsen,~K.~W.  Heats of Formation of Solids with Error
  Estimation: {{The mBEEF}} Functional with and without Fitted Reference
  Energies,  \textit{Phys. Rev. B} \textbf{2015,} \textsl{91,} 235201.

\bibitem{simm2016}
Simm,~G.~N.;\ \ Reiher,~M.  Systematic {{Error Estimation}} for {{Chemical
  Reaction Energies}},  \textit{J. Chem. Theory Comput.} \textbf{2016,}
  \textsl{12,} 2762--2773.

\bibitem{proppe2016}
Proppe,~J.;\ \ Husch,~T.;\ \ Simm,~G.~N.;\ \ Reiher,~M.  Uncertainty
  Quantification for Quantum Chemical Models of Complex Reaction Networks,
  \textit{Faraday Discuss.} \textbf{2016,} \textsl{195,} 497--520.

\bibitem{pernot2017a}
Pernot,~P.  The Parameter Uncertainty Inflation Fallacy,  \textit{J. Chem.
  Phys.} \textbf{2017,} \textsl{147,} 104102.
  
\bibitem{simm2018}
Simm,~G.~N.;\ \ Reiher,~M.  Error-{{Controlled Exploration}} of {{Chemical
  Reaction Networks}} with {{Gaussian Processes}},  \textit{J. Chem. Theory Comput.} \textbf{2018,} \textsl{14,} 5238--5248.

\bibitem{bishop2009}
Bishop,~C.~M. \textit{Pattern Recognition and Machine Learning;} {Springer}:
  New York, 2006.

\bibitem{althorpe2016}
Althorpe,~S.~C. \textit{et al.}\   Fundamentals: General Discussion,
  \textit{Faraday Discuss.} \textbf{2016,} \textsl{195,} 139--169.

\bibitem{althorpe2016a}
Althorpe,~S.~C. \textit{et al.}\   Non-Adiabatic Reactions: General Discussion,
   \textit{Faraday Discuss.} \textbf{2016,} \textsl{195,} 311--344.

\bibitem{angulo2016}
Angulo,~G. \textit{et al.}\   New Methods: General Discussion,  \textit{Faraday
  Discuss.} \textbf{2016,} \textsl{195,} 521--556.

\bibitem{althorpe2016b}
Althorpe,~S. \textit{et al.}\   Application to Large Systems: General
  Discussion,  \textit{Faraday Discuss.} \textbf{2016,} \textsl{195,} 671--698.

\bibitem{turanyi2014}
Tur\'anyi,~T.;\ \ Tomlin,~A.~S. \textit{{Analysis of Kinetic Reaction
  Mechanisms};} {Springer}: Berlin, 2014.

\bibitem{kee1980}
Kee,~R.~J.;\ \ Miller,~J.~A.;\ \ Jefferson,~T.~H. ``{{CHEMKIN}}: A
  General-Purpose, Problem-Independent, Transportable, {{FORTRAN}} Chemical
  Kinetics Code Package'',  Technical Report SAND--80-8003, {Sandia National
  Labs., Livermore CA, United States}, 1980.

\bibitem{kee1989}
Kee,~R.~J.;\ \ Rupley,~F.~M.;\ \ Miller,~J.~A. ``Chemkin-{{II}}: {{A Fortran}}
  Chemical Kinetics Package for the Analysis of Gas-Phase Chemical Kinetics'',
  Technical Report SAND-89-8009, {Sandia National Labs., Livermore CA, United
  States}, 1989.

\bibitem{kee1996}
Kee,~R.~J.;\ \ Rupley,~F.~M.;\ \ Meeks,~E.;\ \ Miller,~J.~A.
  ``{{CHEMKIN}}-{{III}}: {{A FORTRAN}} Chemical Kinetics Package for the
  Analysis of Gas-Phase Chemical and Plasma Kinetics'',  Technical Report
  SAND-96-8216, {Sandia National Labs., Livermore CA, United States}, 1996.

\bibitem{goodwin2009}
Goodwin,~D.~G.;\ \ Moffat,~H.~K.;\ \ Speth,~R.~L. ``Cantera: {{An}}
  Object-Oriented Software Toolkit for Chemical Kinetics, Thermodynamics, and
  Transport Processes'',  Version 2.3.0, DOI: 10.5281/zenodo.170284,
  \url{http://www.cantera.org, 2017}.

\bibitem{hoops2006}
Hoops,~S.;\ \ Sahle,~S.;\ \ Gauges,~R.;\ \ Lee,~C.;\ \ Pahle,~J.;\ \
  Simus,~N.;\ \ Singhal,~M.;\ \ Xu,~L.;\ \ Mendes,~P.;\ \ Kummer,~U.
  {{COPASI}}\textemdash{}a {{COmplex PAthway SImulator}},
  \textit{Bioinformatics} \textbf{2006,} \textsl{22,} 3067--3074.

\bibitem{gillespie2007}
Gillespie,~D.~T.  Stochastic {{Simulation}} of {{Chemical Kinetics}},
  \textit{Annu. Rev. Phys. Chem.} \textbf{2007,} \textsl{58,} 35--55.

\bibitem{glowacki2012}
Glowacki,~D.~R.;\ \ Liang,~C.-H.;\ \ Morley,~C.;\ \ Pilling,~M.~J.;\ \
  Robertson,~S.~H.  {{MESMER}}: {{An Open}}-{{Source Master Equation Solver}}
  for {{Multi}}-{{Energy Well Reactions}},  \textit{J. Phys. Chem. A}
  \textbf{2012,} \textsl{116,} 9545--9560.

\bibitem{shannon2018}
Shannon,~R.;\ \ Glowacki,~D.~R.  A {{Simple}} ``{{Boxed Molecular Kinetics}}''
  {{Approach To Accelerate Rare Events}} in the {{Stochastic Kinetic Master
  Equation}},  \textit{J. Phys. Chem. A} \textbf{2018,} \textsl{122,}
  1531--1541.

\bibitem{glowacki2012a}
Glowacki,~D.~R.;\ \ Lockhart,~J.;\ \ Blitz,~M.~A.;\ \ Klippenstein,~S.~J.;\ \
  Pilling,~M.~J.;\ \ Robertson,~S.~H.;\ \ Seakins,~P.~W.  Interception of
  {{Excited Vibrational Quantum States}} by {{O2}} in {{Atmospheric Association
  Reactions}},  \textit{Science} \textbf{2012,} \textsl{337,} 1066--1069.

\bibitem{glowacki2010}
Glowacki,~D.~R.;\ \ Liang,~C.~H.;\ \ Marsden,~S.~P.;\ \ Harvey,~J.~N.;\ \
  Pilling,~M.~J.  Alkene {{Hydroboration}}: {{Hot Intermediates That React
  While They Are Cooling}},  \textit{J. Am. Chem. Soc.} \textbf{2010,}
  \textsl{132,} 13621--13623.

\bibitem{goldman2011}
Goldman,~L.~M.;\ \ Glowacki,~D.~R.;\ \ Carpenter,~B.~K.  Nonstatistical
  {{Dynamics}} in {{Unlikely Places}}: [1,5] {{Hydrogen Migration}} in
  {{Chemically Activated Cyclopentadiene}},  \textit{J. Am. Chem. Soc.}
  \textbf{2011,} \textsl{133,} 5312--5318.

\bibitem{scine}
``{{SCINE}}. {{Software}} for {{Chemical Interaction Networks}}'',  ETH
  Z\"urich, Z\"urich, Switzerland,
  \url{http://www.reiher.ethz.ch/software/scine.html}.

\bibitem{grimmett1992}
Grimmett,~G.;\ \ Stirzaker,~D. \textit{Probability and {{Random Processes}};}
  {Oxford University Press}: New York, 2nd ed., 1992.

\bibitem{kirk2009}
Kirk,~P. D.~W.;\ \ Stumpf,~M. P.~H.  Gaussian Process Regression Bootstrapping:
  Exploring the Effects of Uncertainty in Time Course Data,
  \textit{Bioinformatics} \textbf{2009,} \textsl{25,} 1300--1306.

\bibitem{sutton2016}
Sutton,~J.~E.;\ \ Guo,~W.;\ \ Katsoulakis,~M.~A.;\ \ Vlachos,~D.~G.  Effects of
  Correlated Parameters and Uncertainty in Electronic-Structure-Based Chemical
  Kinetic Modelling,  \textit{Nat. Chem.} \textbf{2016,} \textsl{8,} 331--337.

\bibitem{ramakrishnan2014}
Ramakrishnan,~R.;\ \ Dral,~P.~O.;\ \ Rupp,~M.;\ \ von Lilienfeld,~O.~A.
  Quantum Chemistry Structures and Properties of 134 Kilo Molecules,
  \textit{Sci. Data} \textbf{2014,} \textsl{1,} 140022.

\bibitem{pernot2017}
Pernot,~P.;\ \ Cailliez,~F.  A Critical Review of Statistical
  Calibration/Prediction Models Handling Data Inconsistency and Model
  Inadequacy,  \textit{AIChE J.} \textbf{2017,} \textsl{63,} 4642--4665.

\bibitem{proppe2017}
Proppe,~J.;\ \ Reiher,~M.  Reliable {{Estimation}} of {{Prediction
  Uncertainty}} for {{Physicochemical Property Models}},  \textit{J. Chem.
  Theory Comput.} \textbf{2017,} \textsl{13,} 3297--3317.

\bibitem{eyring1935}
Eyring,~H.  The {{Activated Complex}} in {{Chemical Reactions}},  \textit{J.
  Chem. Phys.} \textbf{1935,} \textsl{3,} 107--115.

\bibitem{meijer2005}
Meijer,~E.  Matrix Algebra for Higher Order Moments,  \textit{Linear Algebra
  Appl.} \textbf{2005,} \textsl{410,} 112--134.

\bibitem{matlab2018a}
``{{MATLAB}}'',  Release 2018a, The MathWorks, Inc., Natick MA, United States,
  2018.

\bibitem{shampine1997}
Shampine,~L.;\ \ Reichelt,~M.  The {{MATLAB ODE Suite}},  \textit{SIAM J. Sci.
  Comput.} \textbf{1997,} \textsl{18,} 1--22.

\bibitem{morris1991}
Morris,~M.~D.  Factorial {{Sampling Plans}} for {{Preliminary Computational
  Experiments}},  \textit{Technometrics} \textbf{1991,} \textsl{33,} 161--174.

\bibitem{besora2018}
Besora,~M.;\ \ Maseras,~F.  Microkinetic Modeling in Homogeneous Catalysis,
  \textit{WIREs Comput. Mol. Sci.} \textbf{2018,}  e1372, DOI:
  10.1002/wcms.1372.

\bibitem{rasmussen2006}
Rasmussen,~C.~E.;\ \ Williams,~C. K.~I. \textit{{Gaussian Processes for Machine
  Learning};} {The MIT Press}: Cambridge MA, 2006.

\bibitem{susnow1997}
Susnow,~R.~G.;\ \ Dean,~A.~M.;\ \ Green,~W.~H.;\ \ Peczak,~P.;\ \
  Broadbelt,~L.~J.  Rate-{{Based Construction}} of {{Kinetic Models}} for
  {{Complex Systems}},  \textit{J. Phys. Chem. A} \textbf{1997,} \textsl{101,}
  3731--3740.

\bibitem{han2017}
Han,~K.;\ \ Green,~W.~H.;\ \ West,~R.~H.  On-the-Fly Pruning for Rate-Based
  Reaction Mechanism Generation,  \textit{Comput. Chem. Eng.} \textbf{2017,}
  \textsl{100,} 1--8.

\bibitem{song2004}
Song,~J. \textit{Building Robust Chemical Reaction Mechanisms: Next Generation
  of Automatic Model Construction Software,} Doctoral thesis, Massachusetts
  Institute of Technology, 2004.

\bibitem{jalan2013}
Jalan,~A.;\ \ West,~R.~H.;\ \ Green,~W.~H.  An {{Extensible Framework}} for
  {{Capturing Solvent Effects}} in {{Computer Generated Kinetic Models}},
  \textit{J. Phys. Chem. B} \textbf{2013,} \textsl{117,} 2955--2970.

\bibitem{grenda2003}
Grenda,~J.~M.;\ \ Androulakis,~I.~P.;\ \ Dean,~A.~M.;\ \ Green,~W.~H.
  Application of {{Computational Kinetic Mechanism Generation}} to {{Model}}
  the {{Autocatalytic Pyrolysis}} of {{Methane}},  \textit{Ind. Eng. Chem.
  Res.} \textbf{2003,} \textsl{42,} 1000--1010.

\bibitem{horn1990}
Horn,~R.~A.;\ \ Johnson,~C.~R. \textit{Matrix {{Analysis}};} {Cambridge
  University Press}: Cambridge, 1990.

\end{thebibliography}
\end{document}